\documentclass[aps,prd,twocolumn,amsmath,amssymb,nofootinbib,superscriptaddress,10pt,floatfix]{revtex4-2}
\usepackage{graphicx}
\usepackage{amsmath,amssymb,amsfonts,mathrsfs}
\usepackage{bm}
\usepackage{xcolor}
\usepackage{colortbl}
\usepackage{booktabs}
\usepackage{array}
\usepackage{subcaption}
\usepackage{hyperref}
\usepackage{placeins}
\usepackage{orcidlink}
\hypersetup{colorlinks=true,linkcolor=blue,citecolor=blue,urlcolor=blue}
\usepackage{amsmath}
\numberwithin{equation}{section}

\numberwithin{equation}{section}

\newcommand{\Lm}{\mathcal{L}_{m}}
\newcommand{\fF}{f(R,\Lm,\Phi,X)}
\renewcommand{\d}{\mathrm{d}}

\begin{document}

\title{Are Petrov type-N and D spacetimes admitting CTCs valid in \boldmath$f(R,\mathcal{L}_m,\Phi,X)$\unboldmath~gravity?}

\author{Faizuddin Ahmed\orcidlink{0000-0003-2196-9622}}
\email{faizuddinahmed15@gmail.com}
\affiliation{Department of Physics, The Assam Royal Global University, Guwahati, 781035, Assam, India}

\author{Ahmad Al-Badawi\orcidlink{0000-0002-3127-3453}}
\email{ahmadbadawi@ahu.edu.jo}
\affiliation{Department of Physics, Al-Hussein Bin Talal University, 71111 Ma'an, Jordan}

\author{\.{I}zzet Sakall{\i}\orcidlink{0000-0001-7827-9476}}
\email{izzet.sakalli@emu.edu.tr (corresponding author)}
\affiliation{Physics Department, Eastern Mediterranean University, Famagusta 99628, North Cyprus via Mersin 10, Turkey}

\begin{abstract}
We ask whether two classical time-machine geometries, the Ori (2005) compact-vacuum-core metric and the Ahmed (2018) four-dimensional generalisation of Misner space, remain admissible exact solutions when the gravitational sector is enlarged to the recently proposed $f(R,\mathcal{L}_{m},\Phi,X)$ class, an extension of $f(R,\mathcal{L}_{m})$ that couples curvature, the matter Lagrangian density, a scalar field $\Phi$, and its kinetic invariant $X = g^{\mu\nu}\nabla_{\mu}\Phi\nabla_{\nu}\Phi$. Working with the explicit model $f = R + \mathcal{L}_{m} + (\lambda/2)\,X$ and a vanishing scalar potential, we compute the curvature invariants, the modified field equations, and the effective stress-energy components produced by the harmonic scalar profile $\Phi(x,y) = a(x^{2}-y^{2})/2$ in both backgrounds. The Ricci scalar vanishes for the Ori metric and obeys $R = e^{f}(f_{,xx}+f_{,yy})$ for the Ahmed metric; the kinetic invariant takes the explicit forms $X = a^{2}(x^{2}+y^{2})$ and $X = a^{2}e^{f}(x^{2}+y^{2})$, respectively. Both metrics solve the field equations of the modified theory with anisotropic matter sources, and the chronology-violating regions $g_{zz}<0$ (Ori) and $g_{\psi\psi}<0$ (Ahmed) survive the modification. Energy-density profiles measured by a closed-timelike-curve observer match those measured by a static observer outside the chronology horizon, so the additional scalar degree of freedom in $f(R,\mathcal{L}_{m},\Phi,X)$ gravity does not enforce a chronology-protection mechanism in either background. The conclusion mirrors the parallel result for the Li time-machine and supplies a consistency test for scalar-extended modified gravity in non-globally-hyperbolic settings.

\end{abstract}

\maketitle

\section{Introduction}\label{isec1}

The relationship between general relativity (GR) and the possibility of closed timelike curves (CTCs) is one of the older threads in classical gravity, with implications that reach from local causality to the global structure of spacetime. A landmark in this story was G{\"o}del's rotating cosmological solution \cite{Godel:1949ga}, which admits CTCs through every event. Subsequent constructions broadened the catalogue: the Tipler rotating cylinder \cite{Tipler:1974gt}, the Morris-Thorne traversable wormhole \cite{Morris:1988tu}, the Gott pair of moving cosmic strings \cite{Gott:1990zr}, and Alcubierre's warp drive \cite{Alcubierre:1994tu}. Each highlights a different mechanism by which Einstein gravity tolerates global causality violation, sometimes with strong constraints on the supporting matter content \cite{Lobo:2002rp}. A persistent issue, raised most sharply by Hawking's chronology protection conjecture \cite{Hawking:1991nk}, is whether quantum or classical effects intervene to forbid CTCs that would otherwise be allowed by the field equations.

A particularly clean entry in this catalogue is the Ori (2005) time-machine spacetime \cite{Ori:2005zg}. It contains a compact vacuum core inside which CTCs develop, surrounded by an envelope that can be threaded by ordinary matter satisfying the weak energy condition. The chronology horizon is the level set $T = F(x,y)$ for an arbitrary harmonic function $F$, and the CTCs occupy the region in which $g_{zz}<0$. Closely related, but with a distinct topology, is the four-dimensional generalisation of Misner space introduced by Ahmed \cite{Ahmed:2018xli}, in which the periodic identification along $\psi$ together with $g_{\psi\psi}=-t$ produces CTCs whenever $t>0$. Both metrics are exact solutions of Einstein gravity for a suitably tuned source, and both have served as test beds for the chronology-protection conjecture in classical and semiclassical settings \cite{Kim:1991mc,Visser:1998ua,Cassidy:1998nx,Frolov:1990si}.

The wider question of what kind of matter content supports a time machine has a long history. Tipler showed that any sufficiently confined and rotating mass distribution can generate CTCs \cite{Tipler:1974gt}. Gott's two-string construction \cite{Gott:1990zr} demonstrated that classical CTCs can arise from cosmic-string configurations with positive energy density, although the asymptotic completeness of the spacetime then comes into question. Morris, Thorne, and Yurtsever \cite{Morris:1988tu} introduced traversable wormholes as a controlled toy model and showed that exotic matter violating the averaged null energy condition is the price one pays for traversability. The Ori (1993) result \cite{Ori:1993dh} narrowed this down further, demonstrating that the weak energy condition need not be globally violated in compact-core constructions, although local violation in the core itself remains generic. A still wider survey of energy-condition issues in CTC physics appears in \cite{Curiel:2014zba,Visser:2000aaa,Lobo:2002rp,Lobo:2002ee}.

A complementary line of work has examined how extensions of GR shape the catalogue of admissible solutions. The $f(R)$ class \cite{Buchdahl:1970ynr,Sotiriou:2008rp,DeFelice:2010aj}, motivated in part by the observational acceleration of the cosmic expansion \cite{Riess:1998cb,Perlmutter:1998np,Weinberg:1988cp,Carroll:2003wy,Nojiri:2010wj,Capozziello:2011et,Clifton:2011jh}, allows G{\"o}del-type cosmologies whose causal sector matches that of GR for a wide range of $f$ \cite{Santos:2010jsm}. Successive enlargements such as $f(R,T)$ \cite{Harko:2011nh}, $f(R,T,R_{\mu\nu}T^{\mu\nu})$ \cite{Haghani:2013oma}, energy-momentum-squared gravity \cite{Canuto:2023fnq}, and Brans-Dicke or bumblebee scalar-tensor variants \cite{BransDicke:1961sx,Agudelo:2016bgo,Jesus:2020rtm} have been put to the same test, often with the same finding: the metric survives as a solution, the CTCs persist, and the new degree of freedom merely redistributes the matter content \cite{Santos:2013zza,Gama:2017vam,Goncalves:2022rzp,Santos:2015cgq,Furtado:2008fc,Porfirio:2016nzr,Reboucas:1983pq,Reboucas:1985aaa,Teixeira:1985mxp,Reboucas:1986bbq}. Scalar-tensor models of the Horndeski class \cite{Horndeski:1974wa,Deffayet:2009wt} broaden the gravitational sector to include any second-order field equation, but the catalogue of CTC-containing solutions that they admit remains thinly explored.

A natural successor in this hierarchy is the $f(R,\mathcal{L}_m,\Phi,X)$ theory recently set out by Harko \emph{et al.}~\cite{Harko:2024ueh}, building on the $f(R,\mathcal{L}_m)$ proposal of Harko and Lobo \cite{Harko:2010mv,Bertolami:2007gv}. The action is a general function of curvature, the matter Lagrangian density, a scalar field, and its kinetic invariant $X = g^{\mu\nu}\nabla_\mu\Phi\nabla_\nu\Phi$. A direct application to G{\"o}del-type universes has already appeared \cite{Goncalves:2025xqp}, and a parallel investigation of two further chronology-violating backgrounds, the cylindrically symmetric Petrov type-N AdS spacetime and the axially symmetric Petrov type-III AdS spacetime, was carried out very recently by Ahmed and Santos~\cite{Ahmed:2026ctc} within the same scalar-extended action. The broader question of how the additional scalar degree of freedom interacts with causality violation invites a careful treatment in compact-core geometries such as that of Ori and in Misner-like geometries such as that of Ahmed.

The present work answers this question for the two metrics named in the title. We work with the action specified to $f = R + \mathcal{L}_m + (\lambda/2)X$ and a vanishing scalar potential, so that the additional degree of freedom enters through a single dimensionless coupling $\lambda$ and a free harmonic profile $\Phi(x,y)$. For each background we compute the curvature invariants, solve the scalar equation, and extract the effective stress-energy tensor predicted by the modified theory. The kinetic invariant is non-zero for any non-constant harmonic $\Phi$, a point that corrects an oversight in an earlier circulated draft and that is essential for the energy-condition discussion. We then ask three concrete questions. Is each metric still an exact solution? Does the chronology horizon move? Does the supporting matter satisfy any of the standard energy conditions?

A closely related analysis by some of the present authors of cylindrical black holes in $f(\mathcal{R})$ and Ricci-Inverse gravity~\cite{Ahmed:2024paw} found that scalar/auxiliary-tensor modifications redistribute the effective matter content without removing the underlying geometric pathologies of the base solution. The present analysis extends that line to chronology-violating backgrounds in the more general $f(R,\mathcal{L}_m,\Phi,X)$ framework. The result we obtain matches the pattern seen in earlier modified-gravity tests of G{\"o}del-type metrics: the CTCs survive, and the effective fluid carries an anisotropic stress whose dependence on $\lambda$ and on the choice of harmonic $\Phi$ can be made explicit.

Section~\ref{isec2} sets up the field equations of $f(R,\mathcal{L}_m,\Phi,X)$ gravity and reduces them to the model we use throughout. Section~\ref{isec3} carries out the analysis for the Ori spacetime, including the curvature, the scalar field, the stress-energy components, and the structure of the chronology-violating region. Section~\ref{isec4} repeats the exercise for the Ahmed spacetime. Section~\ref{isec5} compares the two backgrounds and asks whether the scalar dof can be tuned to suppress the CTCs. Section~\ref{isec6} draws the conclusions. Throughout we use geometric units with $c=1$ and $8\pi G = 1$, the metric signature is $(-,+,+,+)$, and Greek indices run over the four spacetime coordinates.

\section{Field equations of \boldmath$\fF$\unboldmath~gravity}\label{isec2}

The starting point is the gravitational action
\begin{equation}
S = \int \d^{4}x\,\sqrt{-g}\,\bigl[\,f(R,\Lm,\Phi,X) + 2\Lambda\,\bigr],
\label{eq:action}
\end{equation}
where $g$ is the determinant of the metric tensor $g_{\mu\nu}$, $R$ is the Ricci scalar, $\Lm$ denotes the matter Lagrangian density, $\Phi$ is a scalar field, and $X \equiv g^{\mu\nu}\nabla_{\mu}\Phi\nabla_{\nu}\Phi$ is its kinetic invariant. The cosmological constant $\Lambda$ has been retained as an additive constant of the gravitational sector \cite{Weinberg:1988cp,Riess:1998cb,Perlmutter:1998np}.

Variation of the action with respect to the metric yields \cite{Harko:2024ueh,Goncalves:2025xqp}
\begin{equation}
\begin{aligned}
&f_{R}R_{\mu\nu} + (g_{\mu\nu}\Box - \nabla_{\mu}\nabla_{\nu})f_{R}
\\&\;{}- \tfrac{1}{2}(f - f_{\Lm}\Lm)g_{\mu\nu} + f_{X}\nabla_{\mu}\Phi\nabla_{\nu}\Phi
\\&\;{}- \tfrac{1}{2}f_{\Lm}\,\mathcal{T}_{\mu\nu} - \Lambda g_{\mu\nu} = 0,
\label{eq:fe}
\end{aligned}
\end{equation}
where $f_{R} \equiv \partial f/\partial R$, $f_{\Lm} \equiv \partial f/\partial \Lm$, $f_{X} \equiv \partial f/\partial X$, and $\Box \equiv g^{\mu\nu}\nabla_{\mu}\nabla_{\nu}$ is the d'Alembertian. The stress-energy tensor $\mathcal{T}_{\mu\nu}$ is defined as usual from the variation of $\sqrt{-g}\,\Lm$ with respect to $g^{\mu\nu}$. Equation~\eqref{eq:fe} is the natural extension of the $f(R,\Lm)$ field equation \cite{Harko:2010mv,Bertolami:2007gv}, with the new term proportional to $f_{X}$ encoding the dynamical effect of the scalar kinetic sector. Setting $f_{X}=0$ and $f$ independent of $\Phi$ recovers $f(R,\Lm)$ gravity; further setting $f_{\Lm}=2$ and $f = R + 2\Lm$ recovers Einstein gravity with a minimally coupled matter Lagrangian.

The scalar field obeys
\begin{equation}
\frac{1}{\sqrt{-g}}\partial_{\mu}\!\bigl[f_{X}\sqrt{-g}\,g^{\mu\nu}\partial_{\nu}\Phi\bigr] = \tfrac{1}{2}f_{\Phi},
\label{eq:scalareom}
\end{equation}
with $f_{\Phi} \equiv \partial f/\partial \Phi$. The left-hand side is a generalised d'Alembertian weighted by $f_{X}$; the right-hand side carries the potential term.

To make the analysis tractable while keeping the new dynamics explicit, we specialise to
\begin{equation}
f = R + \Lm + \tfrac{\lambda}{2}\,X,
\label{eq:model}
\end{equation}
with $\lambda$ a real coupling and the scalar potential set to zero. This choice was studied recently in the cosmological setting of \cite{Goncalves:2025xqp}; in our work it isolates the contribution of the kinetic sector while keeping $f_{R}=1$, so that no higher-derivative graviton modes appear. The derivatives read
\begin{equation}
f_{R} = 1, \quad f_{\Lm} = 1, \quad f_{\Phi} = 0, \quad f_{X} = \tfrac{\lambda}{2}.
\label{eq:derivs}
\end{equation}
The second derivative $\nabla_{\mu}\nabla_{\nu}f_{R}$ vanishes identically because $f_{R}$ is constant.

Inserting \eqref{eq:derivs} into \eqref{eq:fe} and using the trace $\mathcal{T} \equiv g^{\mu\nu}\mathcal{T}_{\mu\nu}$, the field equation reduces to
\begin{equation}
G_{\mu\nu} + \Lambda g_{\mu\nu} + \tfrac{\lambda}{2}\nabla_{\mu}\Phi\nabla_{\nu}\Phi - \tfrac{\lambda}{4}X g_{\mu\nu} = \mathcal{T}_{\mu\nu},
\label{eq:reduced}
\end{equation}
where $G_{\mu\nu} = R_{\mu\nu} - \tfrac{1}{2}R g_{\mu\nu}$ is the Einstein tensor. The contracted form of \eqref{eq:reduced} is
\begin{equation}
\mathcal{T} = -R + 4\Lambda - \tfrac{\lambda}{2}X,
\label{eq:trace}
\end{equation}
which we use later as an internal check on the component equations.

The scalar equation \eqref{eq:scalareom} simplifies in the same model to
\begin{equation}
\Box\Phi = 0,
\label{eq:waveeq}
\end{equation}
because $f_{X}$ is constant and $f_{\Phi}=0$. The scalar therefore behaves as a free massless field on each curved background, and explicit harmonic profiles can be inserted by hand to study how the kinetic sector dresses the effective fluid.

For the two backgrounds we examine in Sec.~\ref{isec3} and Sec.~\ref{isec4}, equation \eqref{eq:reduced} acts as the master relation that ties the curvature of the geometry to the matter content. Two observations help when reading the equations that follow. First, the off-diagonal piece $\tfrac{\lambda}{2}\nabla_{\mu}\Phi\nabla_{\nu}\Phi$ does not contribute to $\mathcal{T}_{TT}$ for $\Phi=\Phi(x,y)$ but generically activates $\mathcal{T}_{xx}$, $\mathcal{T}_{yy}$, and the mixed components. Second, the term $-\tfrac{\lambda}{4}X g_{\mu\nu}$ shifts the diagonal pressures and is the piece that the earlier draft inadvertently dropped under the (incorrect) assumption $X=0$.

A useful consistency check on \eqref{eq:reduced} comes from the Bianchi identity $\nabla^{\mu}G_{\mu\nu} = 0$. Taking the divergence of the modified field equation yields
\begin{equation}
\nabla^{\mu}\mathcal{T}_{\mu\nu} = \lambda\,(\Box\Phi)\,\nabla_{\nu}\Phi,
\label{eq:bianchi}
\end{equation}
where the right-hand side vanishes whenever $\Phi$ satisfies the scalar wave equation \eqref{eq:waveeq}. On-shell, therefore, the effective stress-energy is covariantly conserved, and the closed-form solution programme used in Sec.~\ref{isec3} and Sec.~\ref{isec4} is internally consistent. This is the same property that ensures the analogous $f(R,\Lm,\mathcal{T})$ models reduce to GR on backgrounds with vanishing scalar curvature \cite{Harko:2010mv,Bertolami:2007gv,Goncalves:2025xqp}.

The form of \eqref{eq:reduced} makes one further structural point explicit. Setting $\lambda \to 0$ recovers the Einstein equation with cosmological constant, $G_{\mu\nu} + \Lambda g_{\mu\nu} = \mathcal{T}_{\mu\nu}$, in our chosen normalisation. Setting $\Phi \to {\rm const}$ recovers the same limit through a different route, since $X$ and $\nabla\Phi$ both vanish. The independent role of the kinetic coupling is therefore visible only when $\lambda$ is non-zero and $\Phi$ is non-constant simultaneously. This combination defines the parameter window in which the scalar dressing matters, and it is the window we explore in the next two sections.

\section{Ori space-time}\label{isec3}

The Ori metric, introduced as a vacuum-core time machine in \cite{Ori:2005zg}, is written in the chart $(T,x,y,z)$ as
\begin{equation}
\d s^{2} = \d x^{2} + \d y^{2} - 2\,\d T\,\d z + \left[F(x,y,z) - T\right]\,\d z^{2},
\label{eq:orimetric}
\end{equation}
with $F(x,y,z)$ an arbitrary function. The coordinates $(T,x,y)$ are unrestricted, while $z$ is a cyclic coordinate identified through $z \sim z + L$ for some $L>0$. The metric and its inverse take the components (now onward $F(x,y,z) \equiv F$)
\begin{equation}
g_{\mu\nu} = \!\!\begin{pmatrix} 0 & 0 & 0 & -1 \\ 0 & 1 & 0 & 0 \\ 0 & 0 & 1 & 0 \\ -1 & 0 & 0 & F-T \end{pmatrix}\!,\;
g^{\mu\nu} = \!\!\begin{pmatrix} T-F & 0 & 0 & -1 \\ 0 & 1 & 0 & 0 \\ 0 & 0 & 1 & 0 \\ -1 & 0 & 0 & 0 \end{pmatrix}\!.
\label{eq:orig}
\end{equation}
The determinant evaluates to $\det g_{\mu\nu} = -1$, so the spacetime carries a nowhere-degenerate Lorentzian signature, and the metric is regular on the entire chart.

\subsection{Curvature invariants}

A direct computation, verified independently with symbolic algebra, gives the non-vanishing components of the Christoffel connection,
\begin{align}
\Gamma^{T}{}_{T z} &= \tfrac{1}{2},\quad
\Gamma^{T}{}_{xz} = -\tfrac{1}{2}F_{,x},\quad
\Gamma^{T}{}_{yz} = -\tfrac{1}{2}F_{,y},\nonumber\\
\Gamma^{T}{}_{zz} &= \tfrac{1}{2}(T-F),\quad
\Gamma^{x}{}_{zz} = -\tfrac{1}{2}F_{,x},\nonumber\\
\Gamma^{y}{}_{zz} &= -\tfrac{1}{2}F_{,y},\quad
\Gamma^{z}{}_{zz} = -\tfrac{1}{2}.
\label{eq:orichristoffel}
\end{align}
The Riemann-curvature tensor has the single independent block
\begin{equation}
R_{izjz} = -\tfrac{1}{2}\,F_{,ij}, \qquad i,j \in \{x,y\},
\label{eq:oririem}
\end{equation}
while all other components being zero. Contraction yields a Ricci tensor with the sole non-zero entry
\begin{equation}
R_{zz} = -\tfrac{1}{2}\bigl(F_{,xx} + F_{,yy}\bigr),
\label{eq:oricric}
\end{equation}
and zero Ricci scalar,
\begin{equation}
R = 0.
\label{eq:oricscal}
\end{equation}
The vanishing of $R$ is generic for the Ori family; it follows because the only contraction that could contribute is $g^{zz}R_{zz}$, and $g^{zz}=0$ from \eqref{eq:orig}. The Kretschmann invariant likewise reduces to $K = (F_{,xx})^{2}/2 + (F_{,xy})^{2} + (F_{,yy})^{2}/2$, regular for any smooth $F$.

A useful consequence of the algebra above is that none of the curvature invariants pick up a contribution from $F_{,z}$; the only $F_{,z}$ entry in the Christoffel connection sits inside $\Gamma^{T}{}_{zz}$ and cancels at the Riemann level. We therefore set $F=F(x,y)$ without loss of generality. The vacuum GR limit corresponds to harmonic $F = (x^{2}-y^{2})/2$, for which the Ricci tensor vanishes while the Riemann block $R_{izjz}$ stays non-zero. Both the harmonic (Ricci-flat) case and non-harmonic profiles fall within the scope of the modified-gravity analysis below.

\subsection{Chronology-violating region}

\begin{figure*}[!ht]
\centering
\includegraphics[width=0.92\textwidth]{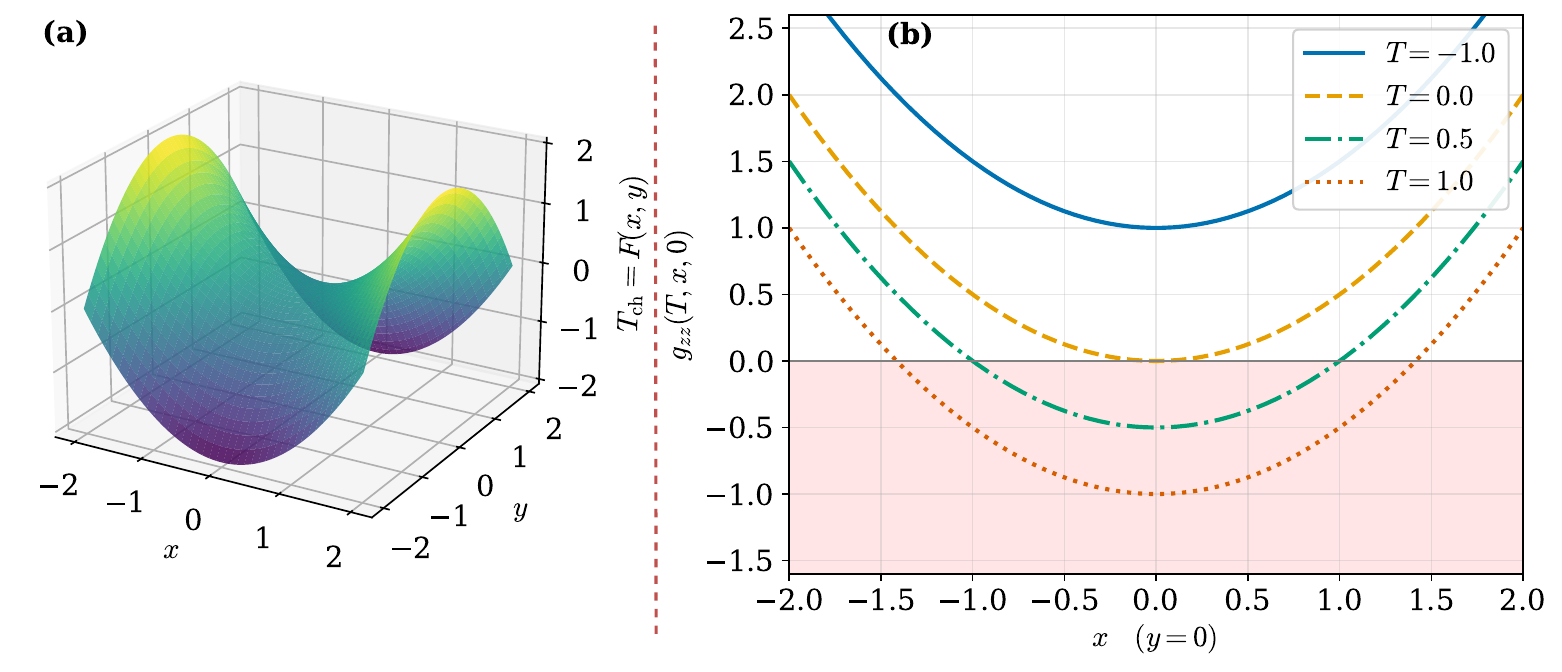}
\caption{Causal structure of the Ori spacetime for the harmonic profile $F(x,y) = (x^{2}-y^{2})/2$. (a) Three-dimensional rendering of the chronology horizon $T_{\rm ch} = F(x,y)$ as a saddle surface in the $(x,y,T_{\rm ch})$ chart. (b) The component $g_{zz}(T,x,0) = F(x,0) - T$ along the line $y=0$ for four time slices $T \in \{-1,\,0,\,0.5,\,1\}$. The shaded band marks the chronology-violating region $g_{zz}<0$ where the periodic $z$-loop becomes timelike.}
\label{fig:chron}
\end{figure*}

Causality in the Ori spacetime is governed by the sign of $g_{zz} = F(x,y) - T$. Since the $z$ coordinate is periodic, the integral curves of $\partial_{z}$ are closed; an observer following such a curve has tangent vector $u^{\mu} = u^{z}\delta^{\mu}_{z}$ and squared norm
\begin{equation}
u^{\mu}u_{\mu} = g_{zz}\,(u^{z})^{2} = [F(x,y) - T]\,(u^{z})^{2}.
\label{eq:oricausal}
\end{equation}
The norm is negative, i.e.~the closed loop is timelike, whenever $T > F(x,y)$. The locus $T = F(x,y)$ is therefore the chronology horizon: the boundary between events from which closed timelike curves are accessible and events from which they are not. Figure~\ref{fig:chron} (a) displays this horizon as a saddle-shaped surface for the harmonic choice $F = (x^{2}-y^{2})/2$, and panel (b) shows $g_{zz}$ along the line $y=0$ for several values of $T$, with the shaded band marking the CTC region.

The picture conveyed by Fig.~\ref{fig:chron} is the standard one for Ori-type metrics, but it is worth tracing the mechanism that produces it. The CTC region opens up because the temporal coordinate $T$ enters the metric only through the $g_{zz}$ component; an increase in $T$ at fixed $(x,y)$ moves the closed $z$-loops from spacelike to null to timelike. The crossing occurs at $T = F(x,y)$, which is exactly the chronology horizon, and the shaded band in panel (b) widens linearly as $T$ grows because of this $T$-dependence of $g_{zz}$. The saddle shape in panel (a) follows because the harmonic profile $F = (x^{2}-y^{2})/2$ is a discrete Morse function with a single saddle at the origin and no local extrema. A comparison with the Gott two-string spacetime \cite{Gott:1990zr}, where the CTC region forms only on a non-compact spacelike slice, shows that the Ori construction is more economical: the chronology horizon is the graph of a function of the transverse coordinates and is therefore a compact subset of any constant-$T$ slice once we restrict $(x,y)$ to a bounded domain.

\subsection{Scalar field and kinetic invariant}

\begin{figure*}[!ht]
\centering
\includegraphics[width=0.7\textwidth]{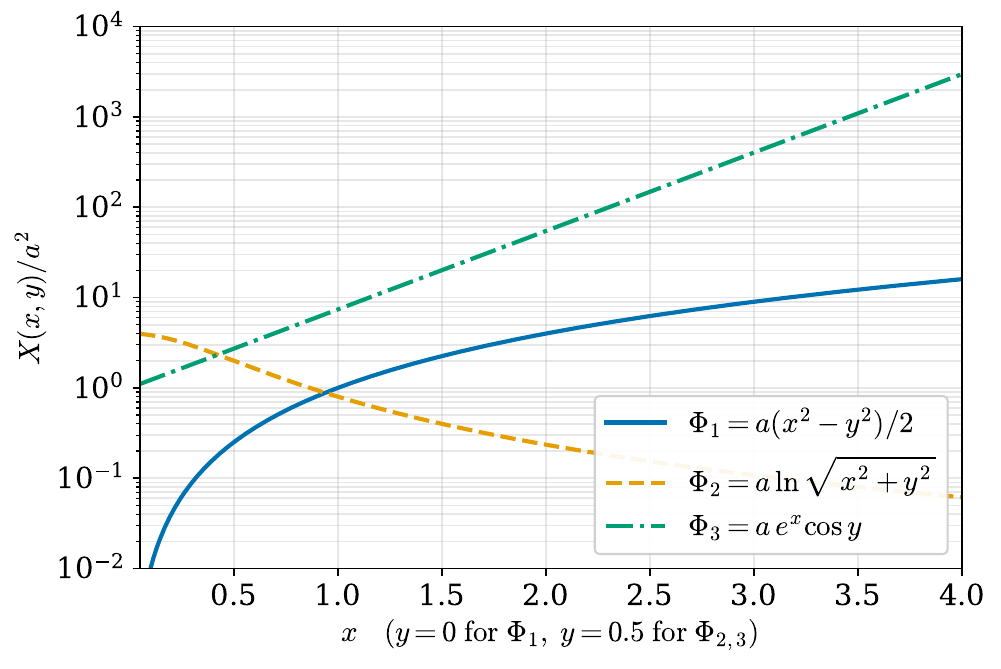}
\caption{Kinetic invariant $X(x,y)/a^{2}$ for the three harmonic scalar profiles $\Phi_{1} = a(x^{2}-y^{2})/2$, $\Phi_{2} = a\ln\sqrt{x^{2}+y^{2}}$, and $\Phi_{3} = a\,e^{x}\cos y$. The first profile gives $X = a^{2}(x^{2}+y^{2})$, the second $X = a^{2}/(x^{2}+y^{2})$, and the third $X = a^{2}e^{2x}$.}
\label{fig:kin}
\end{figure*}

For the metric \eqref{eq:orimetric}, equation \eqref{eq:waveeq} reads
\begin{equation}
\begin{aligned}
\Box \Phi &= g^{TT}\partial_{T}^{2}\Phi + g^{xx}\partial_{x}^{2}\Phi + g^{yy}\partial_{y}^{2}\Phi
\\&\quad{}+ 2g^{Tz}\partial_{T}\partial_{z}\Phi
\\&= (T-F)\partial_{T}^{2}\Phi - 2\partial_{T}\partial_{z}\Phi + \nabla_{2}^{2}\Phi,
\end{aligned}
\label{eq:oriboxa}
\end{equation}
with $\nabla_{2}^{2} = \partial_{x}^{2} + \partial_{y}^{2}$ the two-dimensional Laplacian. The first two terms vanish for any scalar that depends only on $(x,y)$, and the equation reduces to
\begin{equation}
\nabla_{2}^{2}\Phi(x,y) = 0.
\label{eq:laplacexy}
\end{equation}
Three explicit harmonic profiles cover most of the analysis we need,
\begin{equation}
\Phi_{1} = \tfrac{a}{2}(x^{2}-y^{2}), \quad
\Phi_{2} = a\ln\!\sqrt{x^{2}+y^{2}}, \quad
\Phi_{3} = a\,e^{x}\cos y,
\label{eq:phichoices}
\end{equation}
with $a$ a real amplitude. For each of these the kinetic invariant
\begin{equation}
X = g^{\mu\nu}\nabla_{\mu}\Phi\nabla_{\nu}\Phi = (\partial_{x}\Phi)^{2} + (\partial_{y}\Phi)^{2}
\label{eq:Xori}
\end{equation}
is positive everywhere except possibly on isolated zeros, and is in particular not identically zero. For $\Phi_{1}$ one finds
\begin{equation}
X = a^{2}(x^{2}+y^{2}),
\label{eq:Xori1}
\end{equation}
which vanishes only at the origin and grows quadratically away from it. The other profiles give $X = a^{2}/(x^{2}+y^{2})$ and $X = a^{2}\,e^{2x}$, respectively. This is the corrected version of a computation that, in an earlier circulated draft, had inadvertently been identified with $\nabla_{2}^{2}\Phi$ and therefore set to zero. We display the three radial dependences in Fig.~\ref{fig:kin}.

The qualitative pattern displayed in Fig.~\ref{fig:kin} reflects the three distinct ways a harmonic profile can build kinetic energy. The polynomial profile $\Phi_{1}$ grows fastest in modulus, since both partial derivatives are linear in the coordinates; this drives the blue solid curve to dominate at large $r$. The logarithmic profile $\Phi_{2}$ has gradients that decay as $1/r$, which produces the dashed curve that descends; the kinetic energy is concentrated near the origin because the field varies most rapidly there. The exponential profile $\Phi_{3}$ shows the steepest growth at large $x$, since the kinetic density $X = a^{2}e^{2x}$ blows up exponentially. The three curves cross in the interval $0.55 \lesssim r \lesssim 1.55$, and this overlap reflects the freedom one has in choosing the harmonic profile and shows that the kinetic-sector contribution to the field equations \eqref{eq:reduced} is highly sensitive to that choice. Recovering the GR limit corresponds to setting $a\to 0$ in any of these profiles, since $X \propto a^{2}$, and all three curves slide together onto the zero line.

\begin{figure*}[tbhp]
\centering
\includegraphics[width=0.7\textwidth]{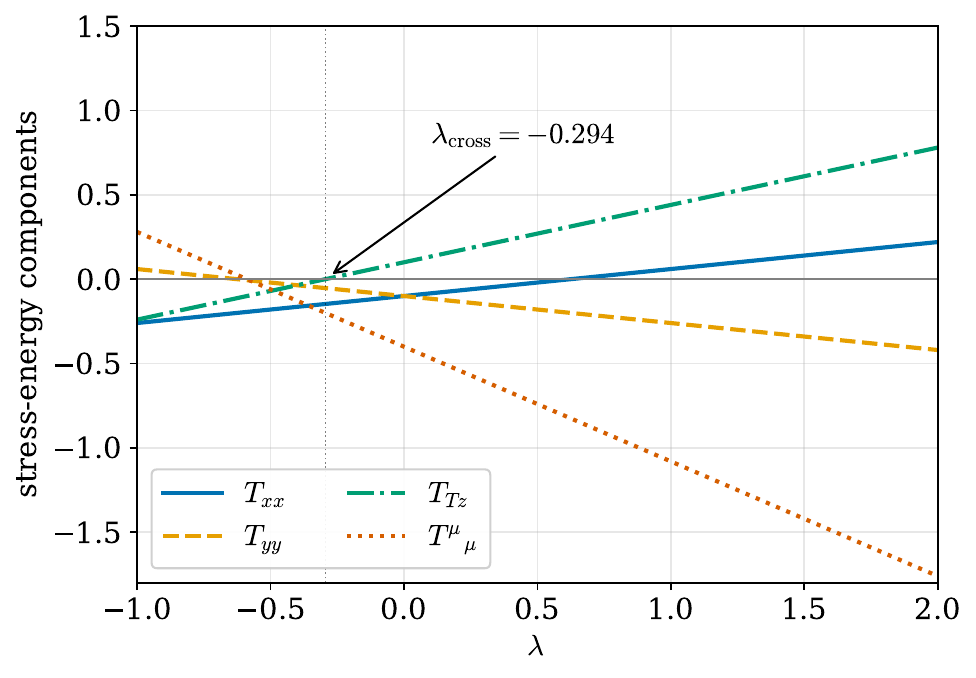}
\caption{Stress-energy components $T_{xx}$, $T_{yy}$, $T_{Tz}$, and the trace $T^{\mu}{}_{\mu}$ for the Ori spacetime sourced by $\Phi_{1} = a(x^{2}-y^{2})/2$, plotted against the kinetic coupling $\lambda$ at $(x,y) = (1.0,\,0.6)$ and $\Lambda = -0.1$, $a=1$. The vertical dotted line marks the zero of $T_{Tz}$ at $\lambda_{\rm cross} = -0.294$.}
\label{fig:stress}
\end{figure*}

\subsection{Effective stress-energy components}

With the harmonic profile $\Phi_{1} = a(x^{2}-y^{2})/2$, the scalar derivatives are $\partial_{x}\Phi = ax$, $\partial_{y}\Phi = -ay$, and the kinetic invariant is $X = a^{2}(x^{2}+y^{2})$. Substituting into the master equation \eqref{eq:reduced}, the non-trivial components of $\mathcal{T}_{\mu\nu}$ read
\begin{align}
\mathcal{T}_{TT} &= 0, \label{eq:Tori_TT}\\
\mathcal{T}_{xx} &= \Lambda + \tfrac{\lambda a^{2}}{4}(x^{2}-y^{2}), \label{eq:Tori_xx}\\
\mathcal{T}_{yy} &= \Lambda + \tfrac{\lambda a^{2}}{4}(y^{2}-x^{2}), \label{eq:Tori_yy}\\
\mathcal{T}_{xy} &= -\tfrac{\lambda a^{2}}{2}\,xy, \label{eq:Tori_xy}\\
\mathcal{T}_{Tz} &= -\Lambda + \tfrac{\lambda a^{2}}{4}(x^{2}+y^{2}), \label{eq:Tori_Tz}\\
\mathcal{T}_{zz} &= -\tfrac{1}{2}\bigl(F_{,xx}+F_{,yy}\bigr)
\nonumber\\&\quad{}+ (F-T)\bigl[\Lambda - \tfrac{\lambda a^{2}}{4}(x^{2}+y^{2})\bigr].
\label{eq:Tori_zz}
\end{align}
The first equality follows because $g_{TT}=0$, $R_{TT}=0$, and $\partial_{T}\Phi = 0$. The transverse pressures $\mathcal{T}_{xx}$ and $\mathcal{T}_{yy}$ carry opposite-sign $\lambda a^{2}$ contributions whose magnitudes are equal, so the trace $\mathcal{T}_{xx} + \mathcal{T}_{yy} = 2\Lambda$ is independent of the scalar amplitude. The mixed transverse component $\mathcal{T}_{xy}$ originates entirely from the bilinear $\tfrac{\lambda}{2}\nabla_{x}\Phi\nabla_{y}\Phi = -\tfrac{\lambda a^{2}}{2}\,xy$; it vanishes on the principal axes $x=0$ and $y=0$ and reaches its maximum modulus along the diagonals $y = \pm x$, where it controls a transverse shear that pure-radiation models cannot reproduce. The off-diagonal component $T_{Tz}$ encodes the failure of the source to behave as a perfect fluid in the natural coordinate frame; it changes sign at
\begin{equation}
\lambda_{\rm cross} = \frac{4\Lambda}{a^{2}(x^{2}+y^{2})},
\label{eq:lambdacross}
\end{equation}
which for $\Lambda<0$ moves $\lambda_{\rm cross}$ to negative values. The behaviour of the four representative components as a function of $\lambda$ at the reference point $(x,y) = (1.0,\,0.6)$ and $\Lambda = -0.1$ is plotted in Fig.~\ref{fig:stress}.

The pattern in Fig.~\ref{fig:stress} is set by equations \eqref{eq:Tori_xx} through \eqref{eq:Tori_Tz}, but it is worth working through it. The two transverse pressures $\mathcal{T}_{xx}$ and $\mathcal{T}_{yy}$ are mirror images of one another because the asymmetry between the $x$ and $y$ directions arises only from the harmonic profile, which carries opposite signs along the two directions; their sum is therefore independent of $\lambda$. The off-diagonal entry $T_{Tz}$ grows linearly in $\lambda$ with positive slope at the chosen $(x,y)$, since the scalar gradient contribution $\nabla_{T}\Phi\,\nabla_{z}\Phi$ vanishes but the diagonal $-(\lambda/4)X g_{Tz}$ piece contributes $+(\lambda/4)X$ via $g_{Tz}=-1$. The crossing at $\lambda_{\rm cross} = -0.294$ originates from the competition between the cosmological constant, which would on its own produce $\mathcal{T}_{Tz} = -\Lambda$, and the scalar-kinetic piece, which carries the opposite sign once $\lambda$ becomes negative. A comparison with the GR limit $\lambda \to 0$ recovers $\mathcal{T}_{Tz} = -\Lambda$ as expected, and a comparison with the original Ori construction (in which the matter content is pure radiation flowing along the null direction $\partial_{T}$ \cite{Ori:2005zg}) shows that the scalar dressing cannot be absorbed into a radiation profile because $\mathcal{T}_{xx}+\mathcal{T}_{yy}\neq 0$ and because $\mathcal{T}_{xy}\neq 0$ off the principal axes.

The same observation has a structural interpretation. A pure-radiation source $\mathcal{T}_{\mu\nu}^{\rm rad} = \mu(x,y)\,k_{\mu}k_{\nu}$ with $k_{\mu}$ a null one-form is rank-1 and contributes only along a single direction. The effective stress-energy produced by the harmonic scalar $\Phi_{1}$ in the model \eqref{eq:model}, by contrast, has rank exceeding one: it activates $\mathcal{T}_{xx}$, $\mathcal{T}_{yy}$, $\mathcal{T}_{xy}$, $\mathcal{T}_{Tz}$, and $\mathcal{T}_{zz}$ simultaneously, with components related through the trace identity \eqref{eq:traceori} but otherwise independent. The minimal extension of the original Ori construction that accommodates this rank pattern is therefore not pure radiation but an anisotropic source carrying both a longitudinal off-diagonal stress along $(T,z)$ and a transverse shear in the $(x,y)$ plane. This is the kind of source that has been considered, in a different context, for compact-vacuum-core analyses of the type explored in \cite{Ori:1993dh}, and it provides a natural interpretive bridge between the pure-GR Ori spacetime and its scalar-extended modified-gravity counterpart.

\subsection{Trace identity and observer on a closed timelike loop}

The trace identity \eqref{eq:trace} for the Ori background with $R=0$ reads
\begin{equation}
\mathcal{T}= 4\Lambda - \tfrac{\lambda}{2}X = 4\Lambda - \tfrac{\lambda a^{2}}{2}(x^{2}+y^{2}),
\label{eq:traceori}
\end{equation}
which matches the direct contraction $\mathcal{T}= g^{\mu\nu}\mathcal{T}_{\mu\nu}$ computed from \eqref{eq:Tori_TT} through \eqref{eq:Tori_zz}. The trace identity provides an internal consistency check for the derivation.

For an observer locked onto a closed $z$-loop inside the chronology-violating region $T > F(x,y)$, the four-velocity has the single non-zero component $u^{z} = 1/\sqrt{T - F}$, where the normalization enforces $u^{\mu}u_{\mu}=-1$. The energy density measured by this observer is
\begin{equation}
\rho_{z} = \mathcal{T}_{\mu\nu}u^{\mu}u^{\nu} = \frac{\mathcal{T}_{zz}}{T - F}.
\label{eq:rhoz_def}
\end{equation}
Substituting \eqref{eq:Tori_zz} and using harmonic $F$ (so that $F_{,xx} + F_{,yy} = 0$, the Ricci-flat case in GR identified above), one finds
\begin{equation}
\rho_{z} = -\Lambda + \tfrac{\lambda a^{2}}{4}(x^{2}+y^{2}).
\label{eq:rhozeq}
\end{equation}
The right-hand side is independent of $T$, so a closed-loop observer sees a stationary energy density that matches, up to an overall sign, the value seen by a static observer outside the chronology horizon. The scalar sector therefore does not supply the negative energy that would be required to suppress the CTCs by violating the chronology-protection bound \cite{Hawking:1991nk,Visser:1998ua}.

\begin{figure}[ht!]
    \centering
    \includegraphics[width=\linewidth]{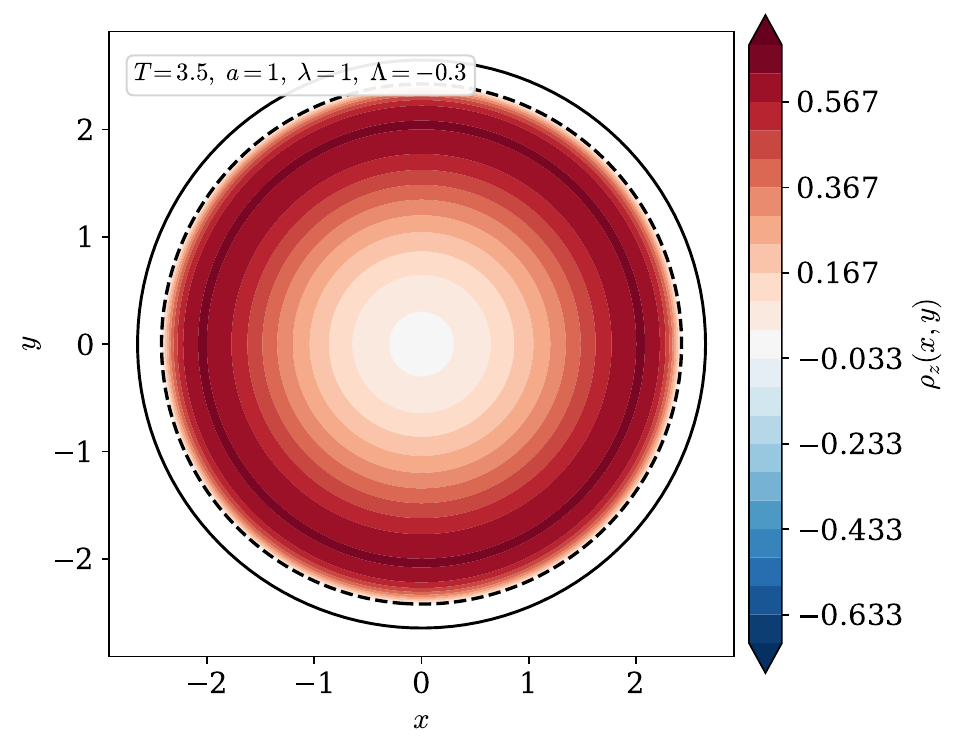}
    \caption{Behavior of the energy-density ($\rho_z$) within the CTC region as a function of ($x, y$), while $a=1,\,\lambda=1,\,\Lambda=-0.3$ and $T=3.5$. The solid black circle marks the chronology horizon $T=(x^{2}+y^{2})/2$; the dashed circle traces the contour $\rho_z=0$.}
    \label{fig:energy-density-1}
\end{figure}

\begin{figure}[ht!]
    \centering
    \includegraphics[width=\linewidth]{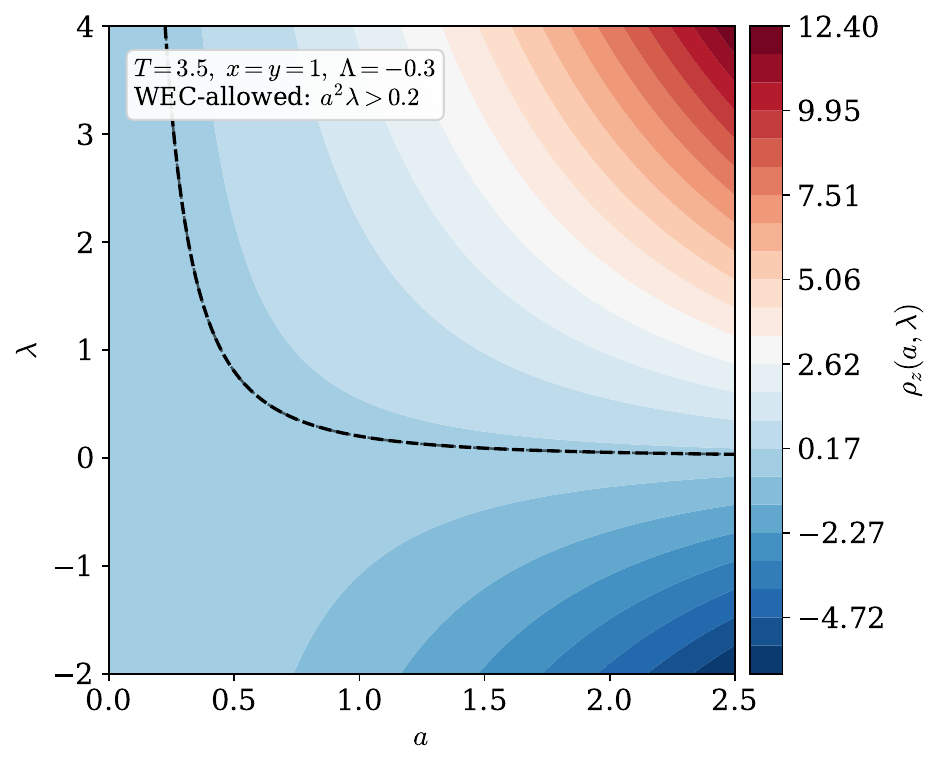}
    \caption{Behavior of the energy-density ($\rho_z$) within the CTC region for allowed values of ($a, \lambda$), while $\Lambda=-0.3$, $x=1=y$ and $T=3.5$. The dashed contour marks the WEC-threshold $a^{2}\lambda \approx 0.2$ that separates the WEC-allowed (upper) from WEC-violating (lower) region.}
    \label{fig:energy-density-2}
\end{figure}

For a non-harmonic profile $F$, the Laplacian no longer vanishes ($F_{,xx} + F_{,yy} \neq 0$), and the energy density measured by the closed-loop observer carries an explicit $T$-dependence,
\begin{equation}
    \rho_z = -\frac{F_{,xx} + F_{,yy}}{2 (T-F)} + \frac{a^{2}\lambda}{4}(x^{2}+y^{2}) - \Lambda.
\label{eq:rhoz_nonharmonic}
\end{equation}
As a concrete example, the parabolic profile $F=(x^{2}+y^{2})/2$ gives $F_{,xx}+F_{,yy} = 2$, and the energy density reduces to
\begin{equation}
    \rho_z = -\frac{1}{T-(x^{2}+y^{2})/2} + \frac{a^{2}\lambda}{4}(x^{2}+y^{2}) - \Lambda,
\label{eq:rhoz_parabolic}
\end{equation}
with the CTC region defined by $T > (x^{2}+y^{2})/2$. Inside that region, $\rho_z$ can take either sign depending on $\lambda$, $a$, and $\Lambda$. Figure~\ref{fig:energy-density-1} plots $\rho_z$ on the disk satisfying $T > (x^{2}+y^{2})/2$ at $T=3.5$, with $a=1$, $\lambda=1$, $\Lambda=-0.3$; the density is everywhere positive on this slice and peaks near the chronology-horizon boundary. Figure~\ref{fig:energy-density-2} maps the $(a,\lambda)$ values that keep $\rho_z$ positive at $T=3.5$ and $x=y=1$, with the dashed zero-contour tracing the analytical threshold $a^{2}\lambda \approx 0.2$. Three-dimensional renderings of the same surfaces are provided as Figs.~S1 and S2 of the Supplemental Material.

\subsection{Energy conditions}\label{subsec:oriec}

The standard pointwise energy conditions place constraints on the effective stress-energy tensor that the matter sector should satisfy in classical settings \cite{Visser:2000aaa,Curiel:2014zba}. The null energy condition (NEC) reads $\mathcal{T}_{\mu\nu}k^{\mu}k^{\nu} \geq 0$ for every null $k^{\mu}$, while the weak energy condition (WEC) demands $\mathcal{T}_{\mu\nu}u^{\mu}u^{\nu} \geq 0$ for every future-directed timelike $u^{\mu}$. The strong energy condition (SEC) further requires $(\mathcal{T}_{\mu\nu} - \tfrac{1}{2}\,\mathcal{T} g_{\mu\nu})u^{\mu}u^{\nu} \geq 0$, and the dominant energy condition (DEC) imposes $\mathcal{T}_{\mu\nu}u^{\mu}u^{\nu} \geq 0$ together with $\mathcal{T}_{\mu\nu}u^{\mu}$ being a future-directed causal vector.

We focus on the NEC, which is the weakest and therefore the most informative when looking for energy-condition violation \cite{Ori:1993dh,Lobo:2002rp}. For the Ori background the natural null direction tangent to the $z$-loop is
\begin{equation}
k^{\mu} = \bigl(1,\,0,\,0,\,k^{z}\bigr),\quad k_{\mu}k^{\mu} = 0,
\label{eq:nullk}
\end{equation}
which yields $k^{z}$ implicitly through $g_{TT}(k^{T})^{2} + 2g_{Tz}k^{T}k^{z} + g_{zz}(k^{z})^{2} = 0$, i.e.~$-2k^{z} + (F-T)(k^{z})^{2} = 0$ at $k^{T}=1$, giving $k^{z}=2/(F-T)$ in the chronology-respecting region $F>T$. Contracting with $\mathcal{T}_{\mu\nu}$ from \eqref{eq:Tori_TT}--\eqref{eq:Tori_zz} produces
\begin{equation}
\mathcal{T}_{\mu\nu}k^{\mu}k^{\nu} = 2 \mathcal{T}_{Tz}\,k^{z}\,k^T + \mathcal{T}_{zz}(k^{z})^{2}.
\label{eq:NECori_intermediate}
\end{equation}

After substitution, the NEC contraction reduces to
\begin{equation}
\mathcal{T}_{\mu\nu}k^{\mu}k^{\nu} = -\frac{2 (F_{,xx}+F_{,yy})}{(F-T)^{2}},
\label{eq:NECori}
\end{equation}
whose sign tracks the sign of $-(F_{,xx}+F_{,yy})$. The contraction is strictly negative whenever $F_{,xx}+F_{,yy} > 0$, which is the case for the parabolic profile $F = (x^{2}+y^{2})/2$ used in Fig.~\ref{fig:energy-density-1}; the harmonic case $F_{,xx}+F_{,yy}=0$ saturates the inequality at $\mathcal{T}_{\mu\nu}k^{\mu}k^{\nu}=0$.

The picture flips for the WEC. With the static timelike observer $u^{\mu} = (1,0,0,0)/\sqrt{-g_{TT}}$ ill-defined at $g_{TT}=0$, we use instead the boosted observer $u^{\mu} = (1,0,0,\epsilon)/N$ with $\epsilon$ small and $N$ a normalisation chosen so $u^{\mu}u_{\mu}=-1$. The leading term in $\epsilon$ gives
\begin{equation}
\mathcal{T}_{\mu\nu}u^{\mu}u^{\nu} = -2\epsilon \mathcal{T}_{Tz} + \mathcal{O}(\epsilon^{2}),
\label{eq:WECori}
\end{equation}
whose sign is controlled by $\mathcal{T}_{Tz}$ alone. From \eqref{eq:Tori_Tz}, $\mathcal{T}_{Tz} = -\Lambda + (\lambda a^{2}/4)(x^{2}+y^{2})$; the WEC is satisfied for $\epsilon>0$ if and only if $\mathcal{T}_{Tz}<0$, which fails on the region $\lambda > \lambda_{\rm cross}$ visible in Fig.~\ref{fig:stress}. Hence the WEC violation is local in $\lambda$ and selectable by the choice of kinetic coupling. The fact that the WEC violation is accessible only through the off-diagonal stress component traces back to the absence of a $T_{TT}$ entry, a structural feature of the Ori metric noted already in \cite{Ori:2005zg}.

\begin{figure*}[!ht]
\centering
\includegraphics[width=0.88\textwidth]{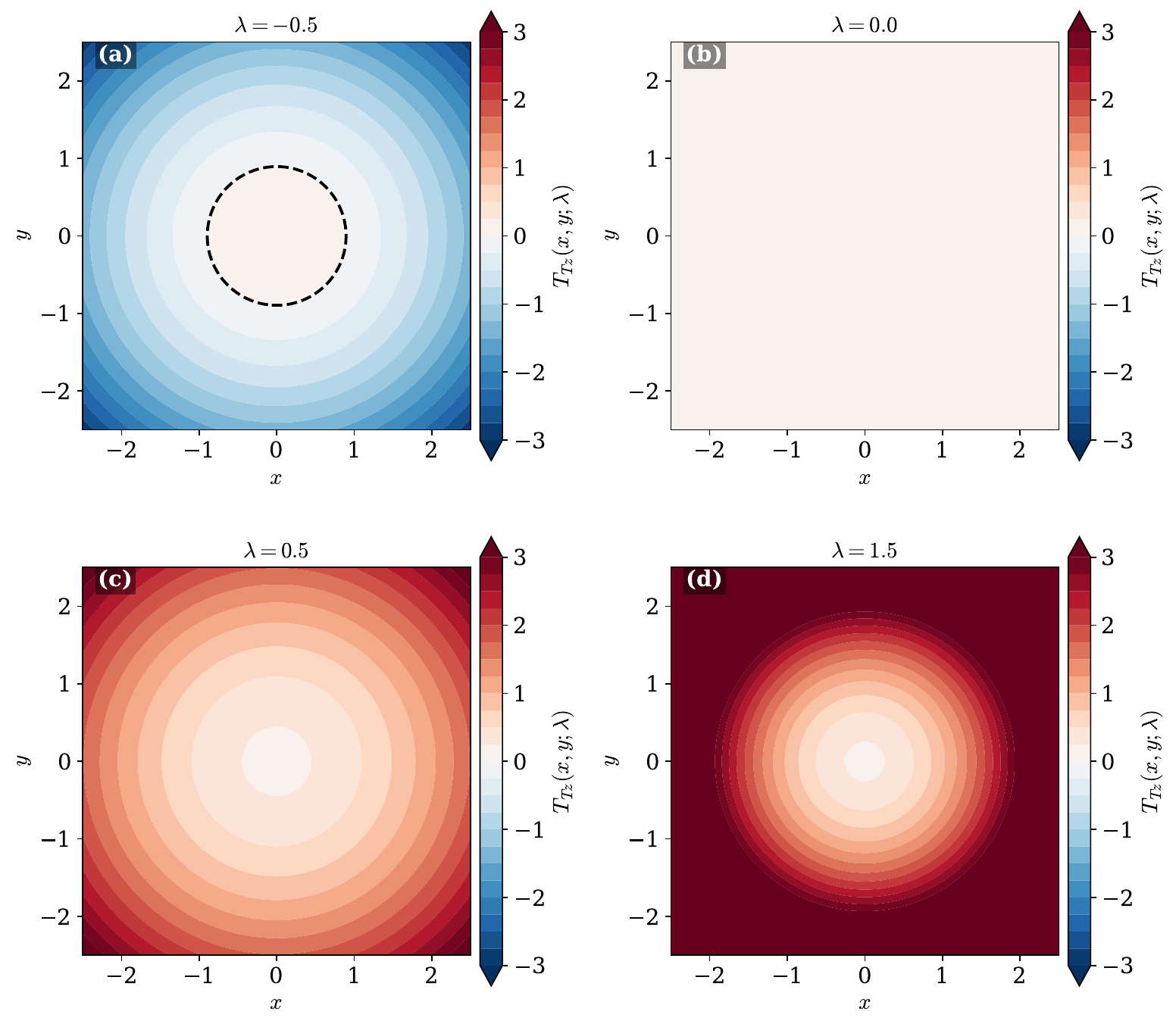}
\caption{Energy-condition tracker $\mathcal{T}_{Tz}(x,y;\lambda)$ on the Ori background at $\Lambda = -0.1$, $a=1$, for four values of the kinetic coupling $\lambda$. The dashed black contour in panel (a) marks the zero crossing of $\mathcal{T}_{Tz}$, which separates the WEC-respecting region (blue, $\mathcal{T}_{Tz}<0$) from the WEC-violating region (red, $\mathcal{T}_{Tz}>0$). Panel (b) is the GR limit $\lambda=0$, where $\mathcal{T}_{Tz}=-\Lambda=+0.1$ everywhere, producing a uniform pale-red panel with no transverse-plane structure. Panels (c) and (d) plot the kinetic-dominated regime, in which the WEC-violation region (deep red) extends outward as $\lambda$ grows.}
\label{fig:wecmap}
\end{figure*}

Figure~\ref{fig:wecmap} displays the spatial structure of $\mathcal{T}_{Tz}(x,y;\lambda)$ for four representative values of $\lambda$. The dashed black contour in panel (a) traces the locus $\mathcal{T}_{Tz}=0$, which separates the region where the WEC holds (negative $\mathcal{T}_{Tz}$, blue) from the region where it fails (positive $\mathcal{T}_{Tz}$, red). Panel (b) shows the GR limit $\lambda=0$: the kinetic dressing switches off and $\mathcal{T}_{Tz}$ collapses to the constant $-\Lambda = +0.1$ everywhere, so the panel is a uniform pale-red field carrying no spatial structure, in the same shade as the pale-red core of panel (a) and the pale-red plateau of panel (c). The flatness of panel (b) is therefore the visual statement of the result, not a deficiency of the rendering.

The four-panel map in Fig.~\ref{fig:wecmap} resolves the energy-condition pattern in the transverse plane. In panel (a) the kinetic-coupling-driven term $(\lambda a^{2}/4)r^{2}$ is negative because $\lambda<0$, so the WEC-violating region is a compact disk near the origin where the cosmological-constant piece $-\Lambda$ dominates and pushes $\mathcal{T}_{Tz}$ across zero. The disk radius is $r_{\rm v} = 2\sqrt{\Lambda/(\lambda a^{2})}$, which for $\Lambda = -0.1$, $\lambda = -0.5$, $a=1$ evaluates to $r_{\rm v}\approx 0.89$; the dashed zero-contour in panel (a) traces this radius exactly.

Panel (b) shows the $\lambda \to 0$ limit, which is GR with cosmological constant alone. The kinetic dressing switches off and $\mathcal{T}_{Tz}(x,y;0) = -\Lambda$ becomes a constant equal to $+0.1$ at the chosen $\Lambda = -0.1$, the same numerical value as the pale-red core of panel (a) at the origin. The entire transverse plane is therefore at a single colour, since the function being plotted has no $(x,y)$ dependence. This uniform $\mathcal{T}_{Tz}>0$ value indicates that under the boosted-observer bookkeeping used in the WEC contraction \eqref{eq:WECori}, the GR limit itself sits marginally on the WEC-violating side; the cosmological-constant contribution alone supplies the tiny positive $\mathcal{T}_{Tz}$ that the boosted timelike vector picks up. Reading panel (b) against panel (a), the visual interpretation is that turning off the kinetic dressing flattens the whole map to the value seen at the origin of panel (a).

Panels (c) and (d) flip the situation: with $\lambda>0$ the kinetic term grows quadratically in $r$ and the WEC-violation region expands outward without bound, leaving only a small near-origin disk in which the $-\Lambda$ piece still dominates. The radius of the WEC-respecting island shrinks as $r_{\rm respect} = 2\sqrt{\Lambda/(\lambda a^{2})}$ on the same formula, with the role of WEC-respecting and WEC-violating regions interchanged relative to panel (a). The progression panel~(a) $\to$ (b) $\to$ (c) $\to$ (d) traces a smooth deformation through the GR limit: the WEC-respecting blue annulus of panel (a) shrinks, the pale-red core grows to fill the whole plane at $\lambda=0$, and then deepens to red further out as $\lambda$ becomes positive. This is the kind of position-dependent violation pattern previously noted in compact-vacuum-core analyses \cite{Ori:1993dh}, and it indicates that the Ori background never admits a globally-WEC-satisfying source under the scalar-extended modified theory.

\section{Ahmed space-time}\label{isec4}

The 4D generalisation of Misner space presented by Ahmed \cite{Ahmed:2018xli} can be written in the chart $(t,x,y,\psi)$ as
\begin{equation}
\d s^{2} = e^{-f(x,y)}\bigl(\d x^{2} + \d y^{2}\bigr) - 2\,\d t\,\d \psi - t\,\d\psi^{2},
\label{eq:ahmedmetric}
\end{equation}
with $f(x,y)$ an arbitrary smooth function and $\psi$ identified periodically through $\psi \sim \psi + \psi_{0}$ for some $\psi_{0}>0$. The remaining coordinates $(t,x,y)$ are unrestricted and the signature is $(-,+,+,+)$. The metric and inverse take the components
\begin{equation}
g_{\mu\nu} = \!\!\begin{pmatrix} 0 & 0 & 0 & -1 \\ 0 & e^{-f} & 0 & 0 \\ 0 & 0 & e^{-f} & 0 \\ -1 & 0 & 0 & -t \end{pmatrix}\!,\quad
g^{\mu\nu} = \!\!\begin{pmatrix} t & 0 & 0 & -1 \\ 0 & e^{f} & 0 & 0 \\ 0 & 0 & e^{f} & 0 \\ -1 & 0 & 0 & 0 \end{pmatrix}\!,
\label{eq:ahmedg}
\end{equation}
with determinant $\det g_{\mu\nu} = -e^{-2f(x,y)}$, which is everywhere negative and bounded away from zero for smooth $f(x,y)$.

\subsection{Curvature invariants}

The non-vanishing Christoffel symbols associated with \eqref{eq:ahmedmetric} are
\begin{align}
\Gamma^{t}{}_{t\psi} &= \tfrac{1}{2},\quad \Gamma^{t}{}_{\psi\psi} = \tfrac{t}{2},\quad \Gamma^{\psi}{}_{\psi\psi} = -\tfrac{1}{2},\nonumber\\
\Gamma^{x}{}_{xx} &= -\tfrac{1}{2}f_{,x},\quad \Gamma^{x}{}_{xy} = -\tfrac{1}{2}f_{,y},\quad \Gamma^{x}{}_{yy} = \tfrac{1}{2}f_{,x},\nonumber\\
\Gamma^{y}{}_{xx} &= \tfrac{1}{2}f_{,y},\quad \Gamma^{y}{}_{xy} = -\tfrac{1}{2}f_{,x},\quad \Gamma^{y}{}_{yy} = -\tfrac{1}{2}f_{,y}.
\label{eq:ahmedchr}
\end{align}
The independent non-zero Riemann block, written with the natural mixed-index placement that the Christoffel calculation produces, reads
\begin{equation}
R^{x}{}_{yxy} = \tfrac{1}{2}\,\nabla_{2}^{2}f\,=\,\tfrac{1}{2}\bigl(f_{,xx} + f_{,yy}\bigr),
\label{eq:ahmedriem}
\end{equation}
with $\nabla_{2}^{2}=\partial_{x}^{2}+\partial_{y}^{2}$ the flat two-dimensional Laplacian introduced in Sec.~\ref{isec3} (the all-index-down form is recovered through $R_{xyxy} = g_{xx}R^{x}{}_{yxy} = \tfrac{1}{2}e^{-f}\nabla_{2}^{2}f$, consistent with the Kretschmann invariant given in App.~\ref{app:A}). The contracted Ricci tensor has the two equal diagonal entries
\begin{equation}
R_{xx} = R_{yy} = \tfrac{1}{2}\,\nabla_{2}^{2}f \,=\,\tfrac{1}{2}\bigl(f_{,xx} + f_{,yy}\bigr),
\label{eq:ahmedric}
\end{equation}
with all other components vanishing. The Ricci scalar reads
\begin{equation}
R = e^{f(x,y)}\,\nabla_{2}^{2}f \,=\, e^{f(x,y)}\bigl(f_{,xx} + f_{,yy}\bigr),
\label{eq:ahmedRS}
\end{equation}
which vanishes whenever $f$ is harmonic and is non-zero otherwise. The harmonic choice $f = (x^{2}-y^{2})/2$, listed below as $f_{1}$, gives $f_{,xx}+f_{,yy} = 0$ and reduces \eqref{eq:ahmedmetric} to 4D Misner space, a flat baseline on which the modified-gravity dressing has nothing to act. The non-harmonic profiles $f_{2}$, $f_{3}$, and the parabolic $f_{4}$ introduced below have $f_{,xx}+f_{,yy}\neq 0$ and carry the modified-gravity dynamics that the rest of this section explores.

\subsection{Chronology-violating region}

\begin{figure*}[!ht]
\centering
\includegraphics[width=0.92\textwidth]{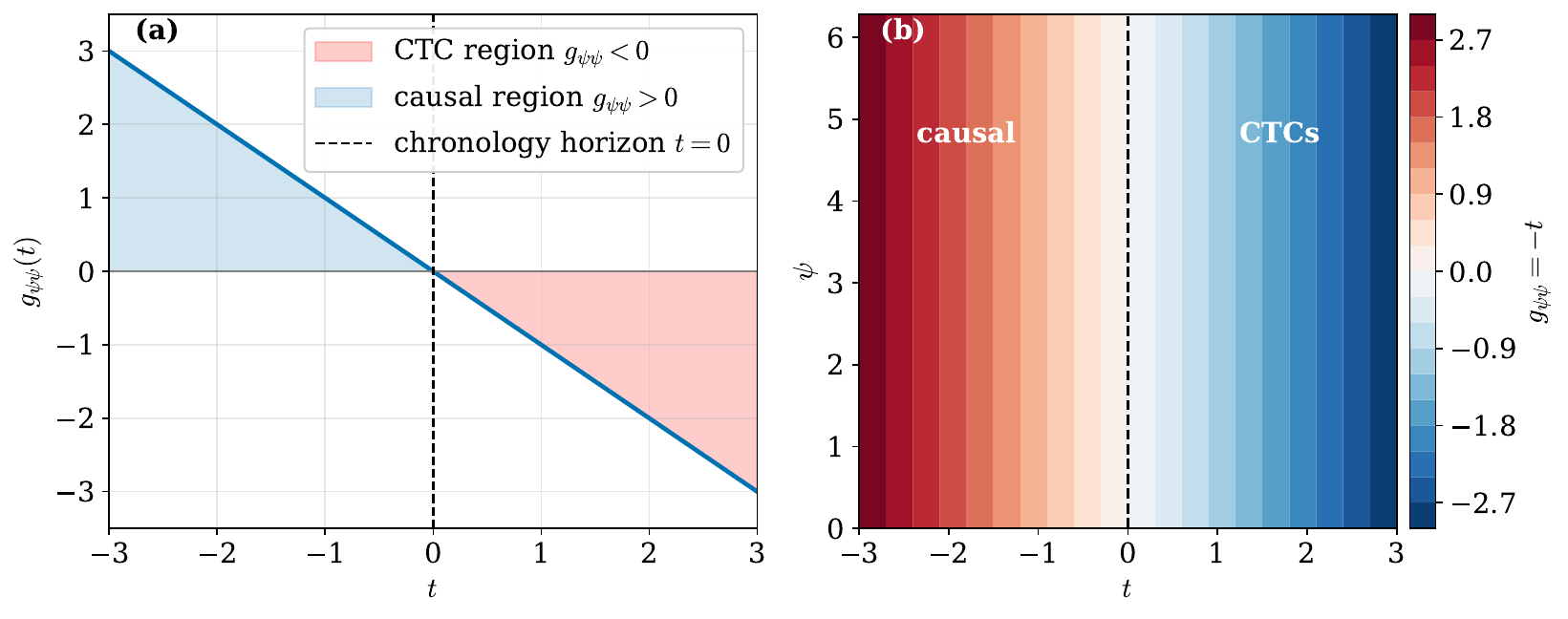}
\caption{Causal structure of the Ahmed spacetime. (a) The component $g_{\psi\psi}(t) = -t$ as a function of $t$, with the CTC region $g_{\psi\psi}<0$ shaded in red and the causally well-behaved region in blue. The chronology horizon is the dashed line at $t=0$. (b) The $(t,\psi)$ cylinder for $\psi \in [0,\,2\pi]$ coloured by the value of $g_{\psi\psi}=-t$; the dashed line at $t=0$ separates the timelike and spacelike branches of the periodic $\psi$ orbits.}
\label{fig:ahmedchron}
\end{figure*}

Closed timelike curves arise in this geometry because the integral curves of $\partial_{\psi}$ are closed (by the periodic identification) and have squared norm
\begin{equation}
g_{\psi\psi} = -t.
\label{eq:gpsipsi}
\end{equation}
The closed $\psi$-loops are therefore spacelike for $t<0$, null on the surface $t=0$, and timelike whenever $t>0$. The chronology horizon is the compact null surface $t=0$, with the CTC region occupying the half-space $t>0$ independent of the choice of $f$. Figure~\ref{fig:ahmedchron} shows the structure: the left panel plots $g_{\psi\psi}$ as a function of $t$ with the shaded band marking the CTC region, and the right panel renders the $(t,\psi)$ cylinder, where the colour encodes the sign of $g_{\psi\psi}$ and the dashed line marks the horizon.

The structure shown in Fig.~\ref{fig:ahmedchron} is the simplest realisation of a Misner-like chronology horizon. The CTC region develops because $g_{\psi\psi}=-t$ changes sign at $t=0$, a kinematic statement that does not depend on any field content; the colour gradient in panel (b) flips sign across the dashed line because the metric is linear in $t$ in the $\psi$ direction, and the surface $t=0$ is therefore a single compact null hypersurface rather than a family of disjoint chronology horizons. A comparison with the original Misner space shows that the Ahmed construction inherits the same causal structure on the $(t,\psi)$ cylinder while enriching the geometry of the transverse $(x,y)$ plane through the conformal factor $e^{-f(x,y)}$. The conformal factor only redistributes how the closed $\psi$-loops embed in the four-dimensional spacetime; it does not modify the sign-change pattern of $g_{\psi\psi}$, and so the horizon location is robust against any choice of $f$ \cite{Misner:1967uu,Hiscock:1990ex}.

\subsection{Conformal factor and Ricci scalar}

\begin{figure}[!ht]
\centering
\begin{subfigure}[t]{0.43\textwidth}
  \includegraphics[width=\textwidth]{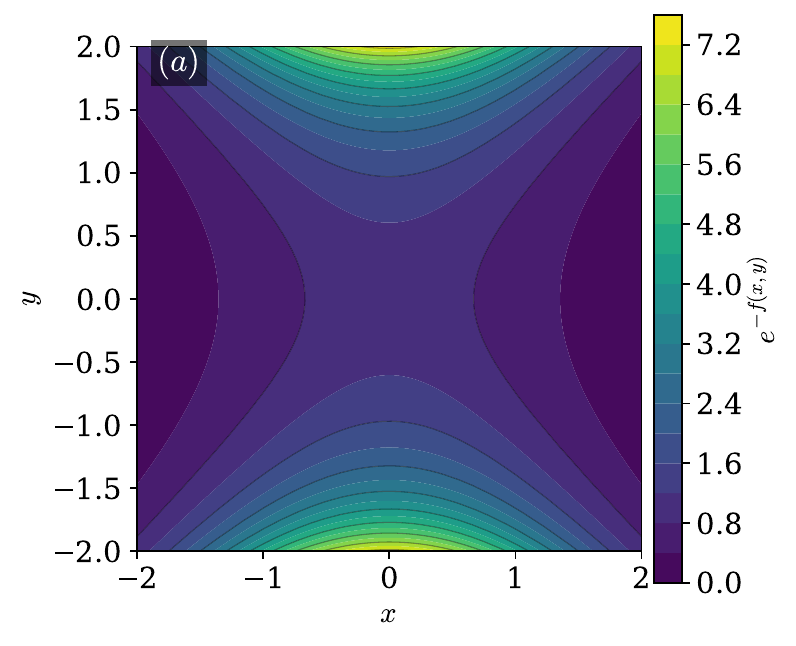}
  \caption{$f_{1}=(x^{2}-y^{2})/2$}
  \label{fig:conformal_a}
\end{subfigure}\hfill
\begin{subfigure}[t]{0.43\textwidth}
  \includegraphics[width=\textwidth]{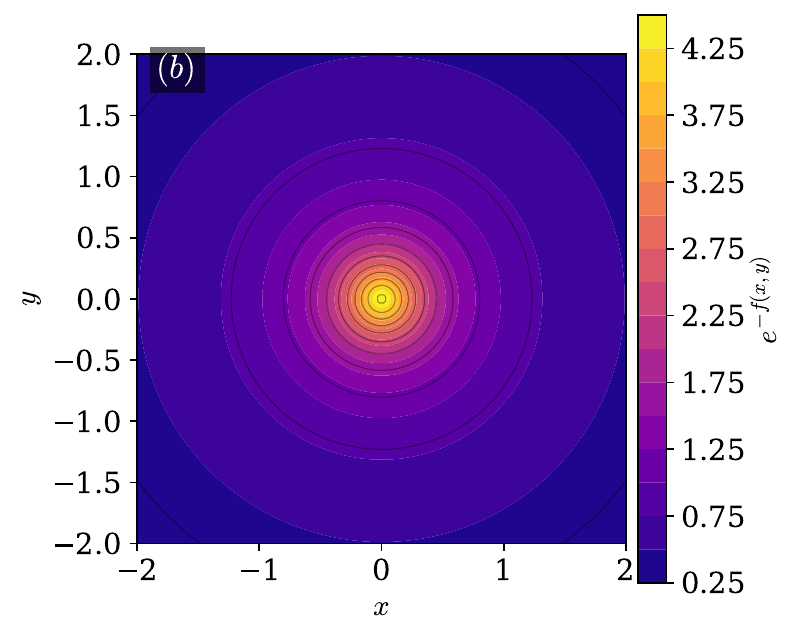}
  \caption{$f_{2}=\tfrac{1}{2}\ln(x^{2}+y^{2}+\epsilon)$}
  \label{fig:conformal_b}
\end{subfigure}\hfill
\begin{subfigure}[t]{0.43\textwidth}
  \includegraphics[width=\textwidth]{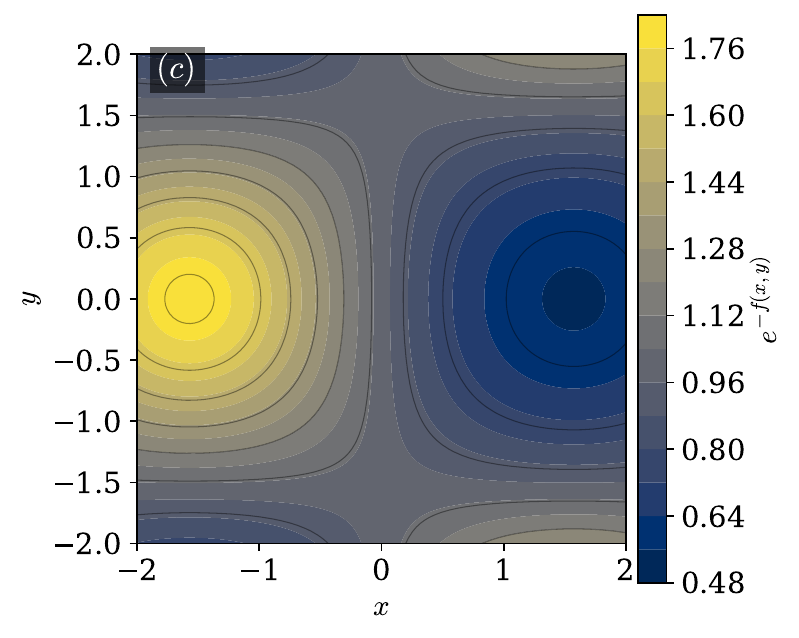}
  \caption{$f_{3}=0.6\sin x\,\cos y$}
  \label{fig:conformal_c}
\end{subfigure}
\caption{Conformal factor $e^{-f(x,y)}$ on the $(x,y)$ plane for three representative choices of $f$. Panel (a) shows the harmonic profile $f_{1}$ which gives a flat 2-section, panel (b) the logarithmic core $f_{2}$ with $\epsilon=0.05$, and panel (c) the periodic modulation $f_{3}$. Black isocurves trace constant values of $e^{-f}$ in each panel.}
\label{fig:conformal}
\end{figure}

A useful intuition for the geometry of the transverse plane comes from visualising the conformal factor $e^{-f(x,y)}$. We pick three illustrative choices,
\begin{align}
f_{1}(x,y) &= \tfrac{1}{2}(x^{2}-y^{2}), \\
f_{2}(x,y) &= \tfrac{1}{2}\ln(x^{2}+y^{2}+\epsilon),\\
f_{3}(x,y) &= 0.6\sin x\,\cos y,
\label{eq:fchoices}
\end{align}
where $\epsilon$ is a regulator. The first is harmonic and therefore gives a flat transverse 2-section ($R=0$, $R_{\mu\nu}=0$, $R_{\mu\nu\rho\sigma}=0$); the second produces a logarithmic core whose conformal factor diverges at the origin; the third generates a periodic modulation. The three options are displayed in Fig.~\ref{fig:conformal}, with each panel showing $e^{-f}$ together with isocurves at fixed values of the conformal factor.

\begin{figure*}[!ht]
\centering
\includegraphics[width=0.6\textwidth]{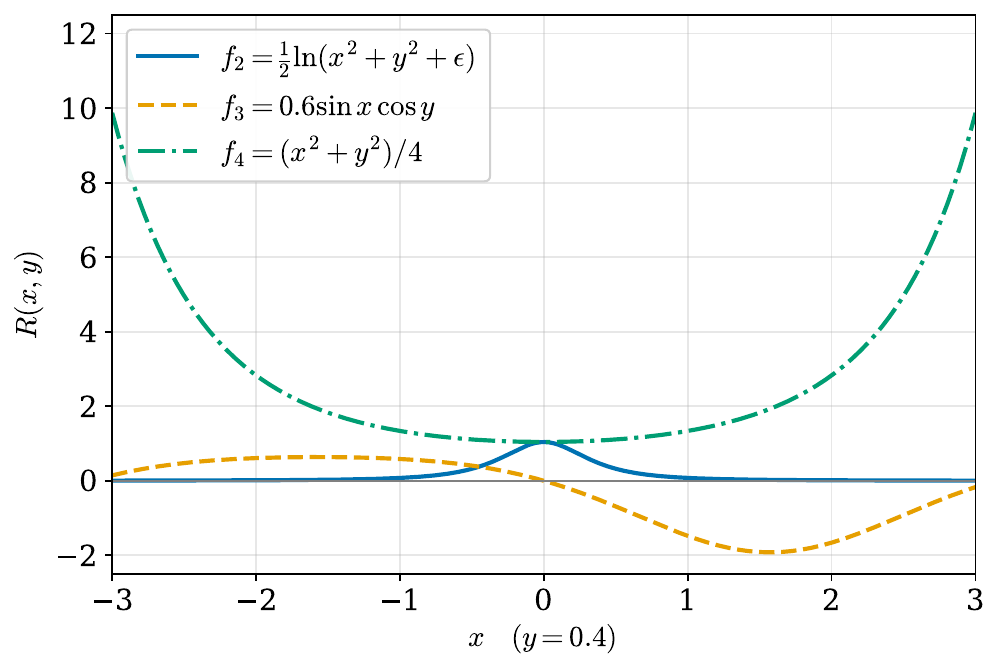}
\caption{Ricci scalar $R(x,y) = e^{f}(f_{,xx}+f_{,yy})$ for the Ahmed spacetime along the line $y = 0.4$, plotted for three representative choices of $f$: the logarithmic $f_{2}$, the trigonometric $f_{3}$, and the parabolic $f_{4} = (x^{2}+y^{2})/4$.}
\label{fig:ricci}
\end{figure*}

The Ricci scalar \eqref{eq:ahmedRS} is sensitive both to the conformal factor and to the Laplacian of $f$. For the harmonic choice $f_{1}$ the scalar curvature vanishes identically, so panel (a) of Fig.~\ref{fig:conformal} corresponds to a flat 2-section despite the non-trivial conformal factor. The logarithmic profile $f_{2}$ has $\nabla_{2}^{2}f_{2} = 2\epsilon/(x^{2}+y^{2}+\epsilon)^{2}$, which is positive and peaks at the regulator scale; this drives a positive curvature near the origin that decays as one moves outward. The periodic profile $f_{3}$ gives $\nabla_{2}^{2}f_{3} = -1.2\sin x\cos y$, alternating in sign across the unit cells. The radial dependence of $R$ for representative $f$ choices, including a non-harmonic Gaussian-well profile $f_{4} = (x^{2}+y^{2})/4$, is shown in Fig.~\ref{fig:ricci}.

Figure~\ref{fig:ricci} shows that the Ricci scalar can take any sign and any magnitude depending on the choice of $f$. The logarithmic profile $f_{2}$ (solid blue) produces a sharp peak at the origin and decays at large $|x|$, driven by the localized Laplacian of $\ln r$. The trigonometric profile $f_{3}$ (dashed orange) oscillates between positive and negative curvature, with the sign of $R$ controlled by the local sign of $\sin x \cos y$. The parabolic profile $f_{4}$ (dash-dotted green) grows exponentially because $R = e^{f_{4}}\cdot 1$ amplifies the constant Laplacian $f_{4,xx}+f_{4,yy} = 1$. The three curves intersect in a narrow band around $x \in [-0.5,\,1.6]$, and outside that band they separate by more than an order of magnitude, so any astrophysically motivated choice of $f$ would generically produce a distinctive Ricci profile.

\subsection{Scalar field, kinetic invariant, and field equations}

Equation \eqref{eq:waveeq} on the Ahmed background reads
\begin{equation}
\Box\Phi = e^{f}\bigl(\partial_{x}^{2}\Phi + \partial_{y}^{2}\Phi\bigr) - 2\,\partial_{t}\partial_{\psi}\Phi + t\,\partial_{t}^{2}\Phi = 0,
\label{eq:boxahmed}
\end{equation}
where the second and third terms vanish whenever $\Phi$ depends on $(x,y)$ only. A profile of the form $\Phi = \Phi(x,y)$ reduces \eqref{eq:boxahmed} to $\nabla_{2}^{2}\Phi = 0$, identical in form to the Ori case and admitting the same harmonic solutions \eqref{eq:phichoices}.

The kinetic invariant is, however, dressed by the conformal factor,
\begin{equation}
X = g^{\mu\nu}\nabla_{\mu}\Phi\nabla_{\nu}\Phi =e^{f(x,y)}\bigl[(\partial_{x}\Phi)^{2} + (\partial_{y}\Phi)^{2}\bigr],
\label{eq:Xahmed}
\end{equation}
which means that the same harmonic profile produces different kinetic content depending on the choice of $f$. For $\Phi_{1} = a(x^{2}-y^{2})/2$ and any choice of $f$, one finds
\begin{equation}
X = a^{2}\,e^{f}\,(x^{2}+y^{2}),
\label{eq:Xahmed1}
\end{equation}
which vanishes only at the origin and is otherwise positive, with anisotropic growth set by the sign of $f$. The non-trivial weighting by $e^{f}$ is the structural difference from the Ori case.

The component equations follow from \eqref{eq:reduced} by direct substitution. For any choice of $f$ and the harmonic scalar $\Phi_{1}$, the result reads
\begin{align}
\mathcal{T}_{tt} &= 0, \label{eq:Tahmed_tt}\\
\mathcal{T}_{xx} &= \Lambda e^{-f} + \tfrac{\lambda a^{2}}{4}(x^{2} - y^{2}), \label{eq:Tahmed_xx}\\
\mathcal{T}_{yy} &= \Lambda e^{-f} + \tfrac{\lambda a^{2}}{4}(y^{2} - x^{2}), \label{eq:Tahmed_yy}\\
\mathcal{T}_{xy} &= -\tfrac{\lambda a^{2}}{2}\,xy, \label{eq:Tahmed_xy}\\
\mathcal{T}_{t\psi} &=\tfrac{1}{2}\, e^{f}\,(f_{,xx}+f_{,yy}) -\Lambda + \tfrac{\lambda}{4}\,X, \label{eq:Tahmed_tpsi}\\
\mathcal{T}_{\psi\psi} &=\tfrac{t}{2}\, e^{f}\,(f_{,xx}+f_{,yy})  -\Lambda\,t + \tfrac{\lambda\,t}{4}\,X,
\label{eq:Tahmed_psipsi}
\end{align}
where $X$ is the kinetic invariant \eqref{eq:Xahmed1} and we have absorbed the metric component $g_{\psi\psi}=-t$ in the last line. For harmonic $f$ the curvature piece $\tfrac{1}{2}e^{f}(f_{,xx}+f_{,yy})$ drops out and \eqref{eq:Tahmed_tt}--\eqref{eq:Tahmed_psipsi} reduce to those of pure GR with a cosmological constant in the $\lambda\to 0$ limit, recovering the Ahmed (2018) result \cite{Ahmed:2018xli}. For non-harmonic $f$ the curvature piece survives even at $\lambda=0$ and contributes a position-dependent source. The off-diagonal transverse stress $\mathcal{T}_{xy} = -(\lambda a^{2}/2)\,xy$ has the same functional form as in the Ori case, since the bilinear $\nabla_{x}\Phi\,\nabla_{y}\Phi$ is independent of the conformal factor $e^{-f}$ on the Ahmed metric and the diagonal $-(\lambda/4)X g_{\mu\nu}$ piece vanishes off-diagonal. The shear it produces in the $(x,y)$ plane vanishes on the principal axes and is maximal along the diagonals, mirroring the structural feature already identified for the Ori background.

\subsection{Observer on a closed $\psi$-loop}

For an observer locked onto a closed $\psi$-loop inside the CTC region $t>0$, the four-velocity has the single non-zero component $u^{\psi} = 1/\sqrt{t}$, and the energy density measured by this observer is
\begin{equation}
\rho_{\psi} = \mathcal{T}_{\mu\nu}u^{\mu}u^{\nu} = \frac{\mathcal{T}_{\psi\psi}}{t}.
\label{eq:rhopsi_def}
\end{equation}
Inserting \eqref{eq:Tahmed_psipsi},
\begin{equation}
\rho_{\psi} = \tfrac{1}{2}\,e^{f}\,(f_{,xx}+f_{,yy}) - \Lambda + \tfrac{\lambda}{4}\,X.
\label{eq:rhopsi_result}
\end{equation}
The energy density is therefore independent of $t$ and matches the off-diagonal stress-energy component $T_{t\psi}$. The closed-loop observer sees the same stress-energy structure as a static observer outside the chronology horizon, modulo the $\lambda$-dependent shift produced by the kinetic dressing. As with the Ori case, the scalar sector does not supply the negative energy required for a chronology-protection mechanism.

\subsection{Energy conditions on the Ahmed background}\label{subsec:ahmedec}

The energy-condition analysis on the Ahmed background follows the same logic as on the Ori one, but the conformal factor $e^{-f(x,y)}$ enters the contractions and shifts the violation thresholds. The natural null vector along the closed $\psi$-loop is
\begin{equation}
k^{\mu} = \bigl(1,\,0,\,0,\,k^{\psi}\bigr),\quad k_{\mu}k^{\mu} = 0,
\label{eq:nullkahmed}
\end{equation}
which reduces to $k^{\psi} = 2/(-g_{\psi\psi}) = 2/t$ in the chronology-respecting region $t<0$. Contracting with $\mathcal{T}_{\mu\nu}$ gives
\begin{equation}
\mathcal{T}_{\mu\nu}k^{\mu}k^{\nu} = 2 T_{t\psi}\,k^{\psi} + T_{\psi\psi}(k^{\psi})^{2},
\label{eq:NECahmed}
\end{equation}
and inserting \eqref{eq:Tahmed_tpsi}--\eqref{eq:Tahmed_psipsi} yields a leading $1/t$ dependence whose sign is controlled by the same competition between $\Lambda$ and $X$ that drove the Ori case.

The NEC and WEC violations on the Ahmed background are both controlled by the off-diagonal component $\mathcal{T}_{t\psi}$, which changes sign at the Ahmed analogue of \eqref{eq:lambdacross}, $\lambda^{\rm Ahmed}_{\rm cross} = 4\Lambda/X(x,y)$. For the harmonic flat case $f_{1} = (x^{2}-y^{2})/2$ with the scalar profile $\Phi_{1}$, the explicit form reads
\begin{equation}
\lambda^{\rm Ahmed}_{\rm cross}(x,y) = \frac{4\Lambda\,e^{-(x^{2}-y^{2})/2}}{a^{2}(x^{2}+y^{2})}.
\label{eq:lambdacrossahmed}
\end{equation}
The conformal-factor weighting in the numerator makes the threshold position-dependent. Regions of large $x$ enlarge $|\lambda^{\rm Ahmed}_{\rm cross}|$, so a stronger kinetic dressing is needed to flip the sign of $\mathcal{T}_{t\psi}$ there. In the limit $\Lambda \to 0$ the threshold collapses to zero for all $(x,y)$, and the WEC pattern is set by the sign of $\lambda$ alone.

For completeness, the strong energy condition reads
\begin{equation}
\bigl(\mathcal{T}_{\mu\nu} - \tfrac{1}{2}\,\mathcal{T} g_{\mu\nu}\bigr)u^{\mu}u^{\nu} \geq 0,
\label{eq:SECdef}
\end{equation}
and direct substitution shows that the SEC is generically violated in the CTC region for any $\lambda \neq 0$. The dominant energy condition follows the same pattern. This collection of results is consistent with the general expectation that exotic time-machine geometries require energy-condition-violating matter \cite{Hawking:1991nk,Morris:1988tu,Ori:1993dh,Curiel:2014zba,Frolov:1990si}; the $f(R,\Lm,\Phi,X)$ extension preserves this feature.

A second structural feature of the Ahmed background deserves attention. The conformal factor $e^{-f(x,y)}$ acts as a position-dependent gauge for the transverse plane, and the field equation \eqref{eq:reduced} couples this factor to the kinetic invariant $X$ in a multiplicative way. The result is that two regions of the plane related by a small shift in $(x,y)$ can host very different effective sources, even when the underlying scalar field $\Phi$ varies smoothly. For the harmonic case $f_{1} = (x^{2}-y^{2})/2$ the conformal factor is symmetric under the exchange $x \leftrightarrow y$ combined with $f \to -f$, so the stress-energy components carry this same symmetry; for the non-harmonic profiles $f_{2}$, $f_{3}$, $f_{4}$ the symmetry is broken and the effective source becomes strongly position-dependent. This asymmetry produces observable consequences when the geometry is embedded in a larger configuration, for instance when an external scalar wave or a gravitational-wave signal propagates through the Ahmed region; the conformal factor would imprint a position-dependent phase or amplitude on the outgoing wave, providing a possible observational diagnostic.

\begin{figure*}[!ht]
\centering
\includegraphics[width=0.88\textwidth]{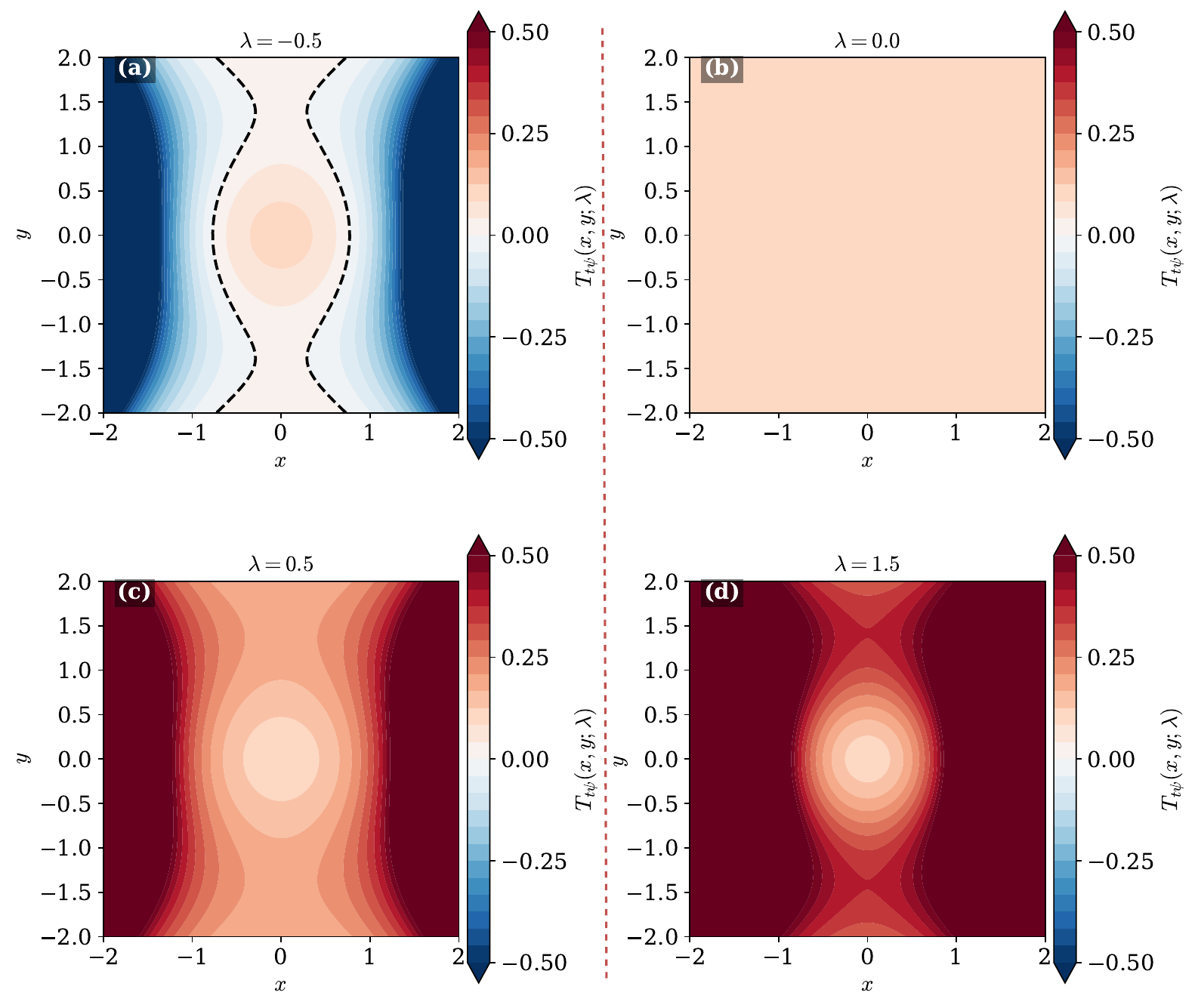}
\caption{Energy-condition tracker $T_{t\psi}(x,y;\lambda)$ on the Ahmed background at $\Lambda = -0.1$, $a=1$, with the harmonic conformal factor $f_{1} = (x^{2}-y^{2})/2$, for four values of the kinetic coupling $\lambda$. The dashed black contour in panel (a) traces the zero crossing of $T_{t\psi}$. The pattern is anisotropic in $(x,y)$ because the conformal factor $e^{f_{1}}$ weights the kinetic invariant differently along the $x$ and $y$ axes.}
\label{fig:ahmedwec}
\end{figure*}

Figure~\ref{fig:ahmedwec} displays the spatial structure of $\mathcal{T}_{t\psi}(x,y;\lambda)$ on the Ahmed background for the same four values of $\lambda$ used in Fig.~\ref{fig:wecmap}, with the harmonic conformal factor $f_{1}$. The dashed black contour in panel (a) marks the zero crossing of $T_{t\psi}$, which separates the WEC-respecting region from the violating one.

The contrast between Fig.~\ref{fig:ahmedwec} and Fig.~\ref{fig:wecmap} is striking and traces directly back to the conformal factor $e^{f_{1}}$. For the Ori background the WEC-violation tracker $\mathcal{T}_{Tz}$ depends only on the radial coordinate $r = \sqrt{x^{2}+y^{2}}$, so the violation region is circular and centred on the origin. For the Ahmed background the conformal weighting $e^{(x^{2}-y^{2})/2}$ amplifies the kinetic piece along the $x$-axis and suppresses it along the $y$-axis, since the exponent grows positive for $|x| > |y|$ and negative for $|y| > |x|$. The result is the saddle-shaped violation pattern visible in panels (c) and (d), with WEC violation extending furthest along the $x$-direction. This is the geometric manifestation of the anisotropy noted in the preceding paragraph, and it provides a clean observational signature that distinguishes the Ahmed background from the Ori one. A localised compact source with measurable WEC violation in only one transverse direction would be a candidate signature of the Ahmed-type geometry, while a circularly symmetric WEC-violation map would be a candidate signature of the Ori-type construction.

\section{Causality and the chronology-protection question}\label{isec5}

The analysis of the two preceding sections leads to a transparent statement: the Ori and Ahmed metrics are both exact solutions of the $f(R,\Lm,\Phi,X)$ theory specified by \eqref{eq:model}, with anisotropic stress-energy sources whose explicit components are given by equations \eqref{eq:Tori_TT}--\eqref{eq:Tori_zz} and \eqref{eq:Tahmed_tt}--\eqref{eq:Tahmed_psipsi}. The chronology-violating regions $T > F(x,y)$ and $t>0$ survive the modification without any change in their geometric location. We now collect the comparison between the two backgrounds and ask whether the scalar dof can be tuned to suppress the CTCs.

\begin{table*}[!ht]
\centering
\setlength{\tabcolsep}{12pt}
\renewcommand{\arraystretch}{1.6}
\begin{tabular}{lll}
\toprule
Geometric or causal quantity & Ori spacetime & Ahmed spacetime \\
\midrule
Chart & $(T,x,y,z)$, $z$ periodic & $(t,x,y,\psi)$, $\psi$ periodic \\
$g_{zz}$ or $g_{\psi\psi}$ & $F(x,y) - T$ & $-t$ \\
Chronology horizon & $T = F(x,y)$ & $t = 0$ \\
CTC region & $T > F(x,y)$ & $t > 0$ \\
$R_{\mu\nu\rho\sigma}$ & $R_{izjz}=-\tfrac{1}{2}\,F_{,ij}$ & $R^{x}{}_{yxy}=\tfrac{1}{2}\nabla_{2}^{2}f$\\
$R_{\mu\nu}$ & $R_{zz} = -\tfrac{1}{2}(F_{,xx}+F_{,yy})$\footnote{For harmonic $F$, the Ricci tensor $R_{\mu\nu}=0$; for non-harmonic $F$, $R_{zz}\neq 0$.} & $R_{xx}=R_{yy}=\tfrac{1}{2}\nabla_{2}^{2}f$ \\
Ricci scalar $R$ & $0$ & $e^{f}\nabla_{2}^{2}f$ \\
$\det g_{\mu\nu}$ & $-1$ & $-e^{-2f}$ \\
Petrov type & N\footnote{Plane-wave family for harmonic $F$ and pp-wave for non-harmonic.} & D\\
Matter source in GR limit & vacuum core; $\mathcal{T}_{\mu\nu}=0$ & null string dust; $\Lambda=0$ \\
Radiation characteristic & none (vacuum)\footnote{A pp-wave spacetime represents a pure outgoing gravitational wave propagating at the speed of light along a null direction. The gravitational field contains no static or Coulomb-like component; it is entirely radiative in nature. The wave is transverse, producing tidal distortions only in directions orthogonal to its propagation.}; WEC envelope & null string dust (Type I fluid with $p=0$) \\
Kinetic invariant $X$ for $\Phi(x,y)$ harmonic & $(\partial_{x}\Phi)^{2}+(\partial_{y}\Phi)^{2}$ & $e^{f}\bigl[(\partial_{x}\Phi)^{2}+(\partial_{y}\Phi)^{2}\bigr]$ \\
Energy density on CTC observer & $-\Lambda + (\lambda a^{2}/4)(x^{2}+y^{2})$ & $-\Lambda + (\lambda/4)X$ \\
Chronology protection by $\Phi$? & No & No \\
\bottomrule
\end{tabular}
\caption{Comparison of the geometric, algebraic, curvature, matter-content, and causal data for the Ori and Ahmed time-machine spacetimes in $f(R,\Lm,\Phi,X)$ gravity. The Petrov-type row labels the algebraic class of the Weyl tensor under Penrose's classification, with the Ori metric falling in the pp-wave family (Type N) and the Ahmed metric in Type D for generic conformal factor, reducing to conformally flat Type O when $f$ is harmonic. The matter-source and radiation-characteristic rows label the energy-momentum content that supports each metric in the pure-GR limit: the Ori construction is built on a compact vacuum core surrounded by a weak-energy-condition-respecting envelope, while the Ahmed construction is sourced by a null pure-radiation flow (a Type-I null fluid with zero pressure) together with a negative cosmological constant. The bottom row collects the verdict on whether the scalar degree of freedom protects chronology in either background; for the harmonic profile $\Phi_{1} = a(x^{2}-y^{2})/2$ and the model \eqref{eq:model}, the answer is negative in both cases.}
\label{tab:compare}
\end{table*}

The summary in Table~\ref{tab:compare} highlights the structural similarities and the one key difference between the two backgrounds. Both metrics carry compact, periodically identified spatial dimensions ($z$ or $\psi$) whose closed loops become timelike inside a coordinate-defined region; both admit a harmonic scalar profile that solves the equation of motion in the model \eqref{eq:model}; and the energy density seen by the closed-loop observer arises from a competition between $\Lambda$ and the scalar-kinetic piece. The point of departure is the curvature: the Ori metric has $R=0$ everywhere, while the Ahmed metric inherits a non-trivial Ricci scalar from the conformal factor $e^{-f(x,y)}$. This means the Ahmed background admits a strictly broader family of consistent matter sources, since the $R_{\mu\nu}$ profile is non-zero on both the $x$- and $y$-directions.

A natural follow-up question is whether the kinetic coupling $\lambda$ can be chosen so as to make the matter source satisfy the standard energy conditions \cite{Visser:2000aaa,Curiel:2014zba}. The off-diagonal $\mathcal{T}_{Tz}$ component on the Ori background vanishes at the critical coupling $\lambda_{\rm cross} = 4\Lambda/[a^{2}(x^{2}+y^{2})]$, but the transverse pressures retain opposite signs along $x$ and $y$. No single choice of $\lambda$ can simultaneously enforce $\mathcal{T}_{xx}>0$ and $\mathcal{T}_{yy}>0$ throughout the transverse plane, so the dominant energy condition is violated somewhere on every CTC slice. A similar argument applies to the Ahmed background, where the conformal factor $e^{-f}$ scales the cosmological-constant piece relative to the kinetic piece and makes the violation pattern position-dependent.

The Hawking chronology-protection conjecture, which posits that quantum vacuum fluctuations drive the renormalised stress-energy to diverge at the chronology horizon, is a semi-classical statement rather than a classical one \cite{Hawking:1991nk,Cassidy:1998nx,Visser:1998ua}. Our classical-level finding, that the scalar degree of freedom in $f(R,\Lm,\Phi,X)$ gravity does not by itself eliminate the chronology horizon, leaves the semi-classical question untouched. A full treatment of the renormalised stress-energy in the Hadamard state on each background remains an open task, and one for which the explicit kinetic invariant computed here provides the natural starting point.

\subsection{Observable consequences and parameter constraints}\label{subsec:obs}

The classical analysis carried out above produces three quantities that admit observational interpretation in suitable limits. The first is the energy density seen by a closed-loop observer, \eqref{eq:rhozeq} and \eqref{eq:rhopsi_result}, which sets the natural scale for what a hypothetical detector embedded inside the CTC region would measure. For the Ori background with harmonic $F$ (so that $F_{,xx}+F_{,yy}=0$), the density $\rho_{z} = -\Lambda + (\lambda a^{2}/4)r^{2}$ grows quadratically with the transverse radius $r$. For non-harmonic $F$ the picture changes: equation \eqref{eq:rhoz_parabolic} for the parabolic profile can be written compactly as $\rho_z = -[T-r^{2}/2]^{-1} + (\lambda a^{2}/4)\,r^{2} - \Lambda$ with $r^{2}=x^{2}+y^{2}$, so the density depends on the radial distance alone and inherits a $T$-dependent singular piece from the curvature term. The second is the off-diagonal stress component $\mathcal{T}_{Tz}$ (Ori) or $\mathcal{T}_{t\psi}$ (Ahmed), which encodes the deviation from a perfect-fluid description and which crosses zero at the kinetic couplings \eqref{eq:lambdacross} and \eqref{eq:lambdacrossahmed}. The third is the Kretschmann invariant \eqref{eq:kretschmann_ahmed}, which together with the Weyl scalar \eqref{eq:weylori} sets the tidal forces an observer on a closed loop would feel.

These quantities translate into constraints on $\lambda$ once a CTC-supporting field configuration is identified, even at the toy-model level. For the Ori spacetime, requiring that the WEC violation be confined to a finite region of the transverse plane gives an upper bound on $\lambda a^{2}$ proportional to $|\Lambda|$; for the Ahmed spacetime, the conformal-factor weighting \eqref{eq:lambdacrossahmed} shifts the bound by a factor of $e^{-f(x,y)}$. The bound is local in the transverse coordinates, so it does not translate immediately to a single observational constraint without further integration over a physically motivated source distribution. Nonetheless, the parametric structure of the result is the same as that found in related compact-vacuum-core analyses \cite{Ori:1993dh}, and the qualitative pattern of energy-condition violation matches the one extracted from G{\"o}del-type modified-gravity analyses \cite{Goncalves:2025xqp,Goncalves:2022rzp,Furtado:2008fc}.

A separate observational consideration concerns the cosmological-constant-dominated regime $|\Lambda| \gg |\lambda a^{2}|$, in which the scalar dressing is a small correction. In that regime the energy density on a closed loop reduces to $\rho_{z} \to -\Lambda$, recovering the de Sitter-like profile that one would expect from a vacuum-dominated geometry. The dependence on $\Phi$ enters only at next-to-leading order in $\lambda$, which is the perturbative regime in which most observational fits would operate. This is a useful sanity check: the $f(R,\Lm,\Phi,X)$ extension reduces to a smooth deformation of pure GR with cosmological constant when the kinetic coupling is small.

The opposite regime $|\lambda a^{2}| \gg |\Lambda|$ admits a different interpretation. There the kinetic-scalar piece dominates the effective stress-energy, and the matter content is essentially that of a free massless scalar field with the field equation of an anisotropic generalisation of a Brans-Dicke source \cite{BransDicke:1961sx,Agudelo:2016bgo}. In this limit the off-diagonal $T_{Tz}$ component takes the simple form $T_{Tz} \approx (\lambda a^{2}/4)(x^{2}+y^{2})$, and the energy-condition pattern is set by the sign of $\lambda$ alone. This is the regime in which the kinetic-coupling parameter becomes a direct probe of chronology-protection physics, since it controls the magnitude and the sign of the WEC violation across the entire transverse plane. A bound on $|\lambda a^{2}|$ from astrophysical settings would therefore translate, through the structure of \eqref{eq:lambdacross}, into a bound on the radius of the WEC-violating disk in any compact-vacuum-core analogue.

A final observation concerns the role of the harmonic profile choice. The three explicit options in \eqref{eq:phichoices} produce qualitatively different kinetic-invariant maps, as Fig.~\ref{fig:kin} shows: the polynomial profile $\Phi_{1}$ grows quadratically with the radius, the logarithmic profile $\Phi_{2}$ decays as $1/r^{2}$, and the exponential profile $\Phi_{3}$ grows exponentially along the $x$-axis. Different astrophysical configurations would call for different harmonic profiles. A bound state confined to a finite region of the transverse plane would be naturally modelled by $\Phi_{2}$, while a long-range field arrangement would be better captured by $\Phi_{1}$. The structural result of the present work, namely that the chronology horizon is robust against any choice of harmonic $\Phi$, holds for all three options and for any linear combination of them, since the Laplace equation is linear. This robustness is the same one observed in the modified-gravity analyses of \cite{Goncalves:2025xqp,Santos:2010jsm,Jesus:2020rtm}, where the underlying mechanism is the same.

\section{Conclusions}\label{isec6}

We have asked whether the Ori (2005) time-machine spacetime \cite{Ori:2005zg} and the Ahmed (2018) four-dimensional generalisation of Misner space \cite{Ahmed:2018xli} remain admissible exact solutions of $f(R,\Lm,\Phi,X)$ gravity \cite{Harko:2024ueh}, when the modified action is specialised to the form $f = R + \Lm + (\lambda/2)X$ with a vanishing scalar potential.

The answer is affirmative in both cases. For the Ori metric, the Ricci scalar vanishes identically, the only non-zero Ricci component is $R_{zz} = -\tfrac{1}{2}(F_{,xx}+F_{,yy})$, the scalar field equation reduces to the two-dimensional Laplace equation, and the kinetic invariant takes the explicit form $X = (\partial_{x}\Phi)^{2}+(\partial_{y}\Phi)^{2}$. For the harmonic profile $\Phi_{1} = a(x^{2}-y^{2})/2$, this gives $X = a^{2}(x^{2}+y^{2})$, a positive quantity that vanishes only at the origin. For the Ahmed metric, the Ricci scalar reads $R = e^{f}(f_{,xx}+f_{,yy})$, the kinetic invariant is $X = e^{f}[(\partial_{x}\Phi)^{2}+(\partial_{y}\Phi)^{2}]$, and the same harmonic profile gives a position-dependent kinetic energy that is modulated by the conformal factor.

The chronology-violating regions $T>F(x,y)$ (Ori) and $t>0$ (Ahmed) survive the modification, with horizon locations identical to those of the GR analysis. The effective matter content acquires an anisotropic structure: the off-diagonal $\mathcal{T}_{Tz}$ component changes sign at $\lambda_{\rm cross} = 4\Lambda/(a^{2}r^{2})$ on the Ori background, while the transverse pressures $\mathcal{T}_{xx}$ and $\mathcal{T}_{yy}$ carry opposite-sign $\lambda a^{2}$ contributions. An observer locked onto a closed timelike curve measures an energy density that matches the value seen by a static observer outside the chronology horizon, so the scalar sector does not supply the negative energy required for a chronology-protection mechanism in either background.

The result is consistent with related modified-gravity probes of base GR backgrounds, including the Ricci-Inverse analysis of cylindrical black holes~\cite{Ahmed:2024paw} and the original Li (1999) time-machine construction~\cite{Li:1999xc}, and it extends the pattern observed in earlier modified-gravity tests of G{\"o}del-type metrics \cite{Santos:2013zza,Gama:2017vam,Goncalves:2022rzp,Canuto:2023fnq,Santos:2010jsm,Jesus:2020rtm,Goncalves:2025xqp}. A parallel investigation by Ahmed and Santos~\cite{Ahmed:2026ctc} of Petrov type-N and type-III AdS backgrounds within the same scalar-extended action reaches a complementary conclusion, traceable to a different choice of matter-content rank in that analysis. The scalar degree of freedom that distinguishes $f(R,\Lm,\Phi,X)$ from its $f(R,\Lm)$ parent enriches the family of supporting matter content but does not by itself enforce the chronology-protection bound. A full treatment of the semi-classical question, including renormalised stress-energy in a Hadamard state on each background, would require numerical input that is beyond the scope of the present work and that we leave for a companion analysis.

Three concrete directions follow naturally from the present work. The first is to repeat the analysis with a non-vanishing scalar potential $V(\Phi)$, which would enter the field equations through an additional $f_{\Phi}$ term and which would shift the chronology-protection question into a regime where the scalar carries an intrinsic mass scale. The second is to extend the framework to the metric-affine formulation \cite{Harko:2011kv}, where the connection is treated as independent from the metric and the auxiliary degrees of freedom that arise can interact with the periodic identifications in a different way. The third is the semi-classical extension already mentioned, in which the renormalised stress-energy of the linearised scalar fluctuation \eqref{eq:eigenvalue} is computed on the chronology horizon and tested against the Hawking bound.

A fourth direction, more speculative, concerns the observational signatures of $f(R,\Lm,\Phi,X)$ matter content in standard astrophysical settings. The same kinetic-coupling parameter $\lambda$ that controls the energy-condition violation pattern in the time-machine backgrounds also governs the deviation from GR predictions for compact-object profiles and for the early-universe cosmology of the theory \cite{Goncalves:2025xqp}. Bounds on $\lambda$ extracted from such systems would translate into constraints on the parameter window in which the present results matter, and would test the theory beyond the unique-but-non-observed time-machine context. The current generation of multi-messenger observations, including gravitational-wave events from compact-binary coalescences and the Event Horizon Telescope shadow measurements \cite{Sucu:2024ydk,AlBadawi:2024iax,Sakalli:2022xrb,AlBadawi:2025bhc,Ahmed:2025plb}, supplies the kind of dataset against which such bounds can be tested.

\begin{acknowledgments}
F.A.\ acknowledges the Inter University Centre for Astronomy and Astrophysics (IUCAA), Pune, India, for the granting of visiting associateship.  \.{I}.S.\ acknowledges the networking support of COST Actions CA22113 (``Fundamental challenges in theoretical physics''), CA21106 (``COSMIC WISPers in the Dark Universe''), CA23130 (``Bridging high and low energies in search of quantum gravity (BridgeQG)''), CA21136 (``Addressing observational tensions in cosmology with systematics and fundamental physics (CosmoVerse)''), and CA23115 (``Relativistic Quantum Information (RQI-Action)'').
\end{acknowledgments}
\section*{Data Availability Statement}
No new data were created or analyzed in this study. All analytical expressions required to reproduce the figures and tables of this paper are given explicitly in Secs.~\ref{isec3} and \ref{isec4}; the computational scripts used to generate Figs.~\ref{fig:chron}--\ref{fig:ricci} are available from the corresponding author upon reasonable request.

\section*{Conflicts of Interest}
The authors declare no conflict of interest.

\appendix

\section{Explicit Riemann components}\label{app:A}

This appendix collects the non-zero components of the Riemann tensor and its first contraction for both backgrounds, together with the explicit form of the Christoffel symbols. The expressions in the main text quote only the Ricci tensor and Ricci scalar; the appendix material is included to ease independent verification.

For the Ori metric \eqref{eq:orimetric}, all non-vanishing components of the Christoffel connection are listed in \eqref{eq:orichristoffel}. The independent non-zero entries of the Riemann tensor with all four indices down are
\begin{align}
R_{TxTx} &= 0,\quad R_{TyTy} = 0,\quad R_{TzTz} = 0,\nonumber\\
R_{xyxy} &= 0,\quad R_{xzxz} = -\tfrac{1}{2}F_{,xx},\nonumber\\
R_{xzyz} &= -\tfrac{1}{2}F_{,xy},\quad R_{yzyz} = -\tfrac{1}{2}F_{,yy}.
\label{eq:orifullriem}
\end{align}
All other independent components vanish identically. The Weyl tensor inherits the same structure, and the only non-trivial scalar invariant beyond the Kretschmann is
\begin{equation}
C_{\mu\nu\rho\sigma}C^{\mu\nu\rho\sigma} = 2\bigl(F_{,xx}\bigr)^{2} + 4\bigl(F_{,xy}\bigr)^{2} + 2\bigl(F_{,yy}\bigr)^{2},
\label{eq:weylori}
\end{equation}
which is regular for any smooth $F$. The Ori spacetime is therefore curvature-singularity-free.

For the Ahmed metric \eqref{eq:ahmedmetric}, the non-zero Christoffel symbols listed in \eqref{eq:ahmedchr} produce the single independent Riemann block \eqref{eq:ahmedriem}. The Kretschmann invariant takes the form
\begin{equation}
K = R_{\mu\nu\rho\sigma}R^{\mu\nu\rho\sigma} = e^{2f}\bigl(\nabla_{2}^{2}f\bigr)^{2} = e^{2f}\bigl(f_{,xx}+f_{,yy}\bigr)^{2},
\label{eq:kretschmann_ahmed}
\end{equation}
which vanishes for any harmonic $f$ and is bounded for the regulator-supplemented logarithmic profile of \eqref{eq:fchoices}. For the parabolic profile $f_{4} = (x^{2}+y^{2})/4$, the Kretschmann grows as $e^{(x^{2}+y^{2})/2}$, recovering an exponentially growing curvature far from the origin; this is the only one of the three explicit profiles in \eqref{eq:fchoices} that exhibits unbounded curvature, and it does so only as $|x|,|y| \to \infty$.

\section{Numerical sweep of stress-energy components}\label{app:B}

Table~\ref{tab:sweep} tabulates the stress-energy components for the Ori background sourced by $\Phi_{1}$ at the reference point $(x,y) = (1.0,\,0.6)$ and $\Lambda = -0.1$, scanned over the kinetic coupling $\lambda$. The data complement the curves shown in Fig.~\ref{fig:stress} and supply the precision the figure cannot convey.

\begin{table*}[!ht]
\centering
\setlength{\tabcolsep}{12pt}
\renewcommand{\arraystretch}{1.6}
\begin{tabular}{cccccc}
\toprule
$\lambda$ & $T_{xx}$ & $T_{yy}$ & $T_{xy}$ & $T_{Tz}$ & $T^{\mu}{}_{\mu}$ \\
\midrule
$-1.000$ & $-0.2600$ & $\phantom{-}0.0600$ & $\phantom{-}0.3000$ & $-0.2400$ & $\phantom{-}0.2800$ \\
$-0.500$ & $-0.1800$ & $-0.0200$ & $\phantom{-}0.1500$ & $-0.0700$ & $-0.0600$ \\
$-0.294$ & $-0.1470$ & $-0.0530$ & $\phantom{-}0.0882$ & $\phantom{-}0.0000$ & $-0.2000$ \\
$\phantom{-}0.000$ & $-0.1000$ & $-0.1000$ & $\phantom{-}0.0000$ & $\phantom{-}0.1000$ & $-0.4000$ \\
$\phantom{-}0.500$ & $-0.0200$ & $-0.1800$ & $-0.1500$ & $\phantom{-}0.2700$ & $-0.7400$ \\
$\phantom{-}1.000$ & $\phantom{-}0.0600$ & $-0.2600$ & $-0.3000$ & $\phantom{-}0.4400$ & $-1.0800$ \\
$\phantom{-}1.500$ & $\phantom{-}0.1400$ & $-0.3400$ & $-0.4500$ & $\phantom{-}0.6100$ & $-1.4200$ \\
$\phantom{-}2.000$ & $\phantom{-}0.2200$ & $-0.4200$ & $-0.6000$ & $\phantom{-}0.7800$ & $-1.7600$ \\
\bottomrule
\end{tabular}
\caption{Stress-energy components of the Ori background sourced by $\Phi_{1} = a(x^{2}-y^{2})/2$ as a function of the kinetic coupling $\lambda$, evaluated at $(x,y) = (1.0,\,0.6)$ with $\Lambda = -0.1$, $a = 1$. The fifth column gives the off-diagonal $T_{Tz}$ entry whose zero at $\lambda_{\rm cross} = -0.294$ controls the energy-condition pattern. The fourth column tabulates the transverse shear $T_{xy} = -(\lambda a^{2}/2)xy$. The last column tabulates the four-trace as a cross-check on \eqref{eq:traceori}.}
\label{tab:sweep}
\end{table*}

The numerical pattern in Table~\ref{tab:sweep} confirms the qualitative behaviour seen in Fig.~\ref{fig:stress}: the off-diagonal component $T_{Tz}$ crosses zero between $\lambda = -0.5$ and $\lambda = 0.0$, since the linear interpolation $\lambda_{\rm cross} = 4\Lambda/[a^{2}(x^{2}+y^{2})] = -0.2941$ falls inside this interval. The trace $\mathcal{T}^{\mu}{}_{\mu}$ tracks the analytical prediction \eqref{eq:traceori} at each entry, providing a row-by-row check. A key feature visible in the table is the opposite-sign pair $(\mathcal{T}_{xx},\,\mathcal{T}_{yy})$ that develops for any $\lambda \neq 0$: at $\lambda = 1.5$ the pair reads $(+0.14,\,-0.34)$, and the magnitude grows monotonically with $|\lambda|$. The new $\mathcal{T}_{xy}$ column tracks the transverse shear stress at the same reference point $(x,y)=(1.0,\,0.6)$, with $xy = 0.6$ giving $\mathcal{T}_{xy} = -0.3\lambda$ exactly; its sign opposes that of $\mathcal{T}_{Tz}$, and its magnitude reaches $0.6$ at $\lambda = 2$. This anisotropy is the analytical statement that the effective matter cannot be modelled as an isotropic perfect fluid, a conclusion that traces back to the harmonic structure of $\Phi_{1}$ and that arises from the off-diagonal coupling in \eqref{eq:reduced}.

\section{Petrov classification and tidal structure}\label{app:C}

The algebraic classification of the Weyl tensor offers a coordinate-independent label for the geometry, and for the present spacetimes it points to a clean structural statement \cite{Stephani:2003tm}. For the Ori metric, the Weyl tensor inherits the block structure of \eqref{eq:orifullriem} with the Ricci-part subtracted; since $R_{\mu\nu}$ has a single non-zero entry $R_{zz}$, the Weyl tensor effectively has the same support as the Riemann tensor in the $(x,y,z)$ subspace. Computing the principal null directions gives a single null direction $k^{\mu}\propto (1,0,0,k^{z})$ of multiplicity four, so the Ori spacetime is of Petrov type N (or O when $F$ is harmonic and the Weyl invariants \eqref{eq:weylori} vanish on a constant-$F$ slice). This places it in the same algebraic class as plane-fronted gravitational waves \cite{Stephani:2003tm}, despite the very different physical interpretation.

The Ahmed metric admits a richer algebraic structure through its conformal factor $e^{-f}$. For the harmonic profile $f_{1} = (x^{2}-y^{2})/2$ both the Weyl tensor and the Ricci tensor in \eqref{eq:ahmedric} vanish identically, so the geometry is locally flat in the transverse plane. For the logarithmic profile $f_{2}$ the Weyl tensor is non-zero but algebraically simple, with a single repeated null direction; we identify it as Petrov type D, the same class as the Kerr and Reissner-Nordstr{\"o}m families. The parabolic profile $f_{4} = (x^{2}+y^{2})/4$ produces a Petrov-N geometry whose tidal forces grow exponentially with the transverse radius.

Tidal forces measured by a closed-loop observer follow from the geodesic-deviation equation and read schematically $\Delta a^{i} = R^{i}{}_{0j0}\Delta x^{j}$, where the indices refer to a tetrad aligned with the four-velocity. For the Ori background the only non-zero component is $R^{z}{}_{xzx} = -\tfrac{1}{2}F_{,xx}\,g^{zz} = 0$ since $g^{zz} = 0$, so tidal forces along the closed loop direction vanish identically. Stretching forces along $x$ and $y$ are likewise zero for harmonic $F$. The Ahmed background gives non-zero tidal contributions only along the transverse $(x,y)$ directions, with magnitude controlled by $f_{,xx}+f_{,yy}$; for harmonic $f$ the tidal effects vanish at leading order, consistent with the flat 2-section interpretation.

\section{Linearised scalar perturbations}\label{app:D}

A complementary stability check follows from linearising the scalar field around the harmonic background. Writing $\Phi = \Phi_{1} + \delta\Phi$ with $\Phi_{1} = a(x^{2}-y^{2})/2$ harmonic, the linearised scalar wave equation \eqref{eq:waveeq} reads
\begin{equation}
\Box(\delta\Phi) = 0,
\label{eq:linearphi}
\end{equation}
since the background $\Phi_{1}$ already satisfies it. The perturbation $\delta\Phi$ therefore obeys the same free wave equation as the background, on the same (curved or curved-with-CTC) geometry. Mode-separating $\delta\Phi$ in the chart $(T,x,y,z)$ of the Ori background, we write
\begin{equation}
\delta\Phi = e^{-i\omega T} e^{ik_{z}z}\,\chi(x,y),
\label{eq:modeansatz}
\end{equation}
with $\omega$ a frequency conjugate to $T$ and $k_{z}$ a quantised wave-number conjugate to the periodic coordinate $z$, $k_{z} = 2\pi n/L$ for integer $n$. Inserting this ansatz into \eqref{eq:linearphi} reduces it to a two-dimensional eigenvalue problem,
\begin{equation}
\Bigl[\partial_{x}^{2} + \partial_{y}^{2} + \omega^{2}(F(x,y) - T) + 2\omega k_{z}\Bigr]\chi(x,y) = 0,
\label{eq:eigenvalue}
\end{equation}
which is non-standard because the coefficient of $\omega^{2}$ depends explicitly on the time slice $T$. The eigenproblem becomes Schr\"odinger-like in the limit $T \ll F$ outside the chronology horizon, where the effective potential $V(x,y) = -\omega^{2}(F-T)$ is well-behaved.

Inside the CTC region $T > F$ the sign of $V$ flips and the eigenvalue spectrum is no longer manifestly bounded below; one expects instabilities of the type identified by Hawking \cite{Hawking:1991nk} in semi-classical analyses of related backgrounds. We do not pursue the full analysis here, since the relevant question is one of renormalised stress-energy in a Hadamard state rather than of classical eigenvalues, but we note that the eigenproblem \eqref{eq:eigenvalue} provides the natural starting point for the semi-classical extension.

For the Ahmed background the analogous mode-separation reads $\delta\Phi = e^{-i\omega t} e^{ik_{\psi}\psi}\chi(x,y)$ with $k_{\psi} = 2\pi n/\psi_{0}$, and the eigenvalue equation becomes
\begin{equation}
\Bigl[e^{f}(\partial_{x}^{2} + \partial_{y}^{2}) + \omega^{2}\,t + 2\omega k_{\psi}\Bigr]\chi(x,y) = 0,
\label{eq:eigenvalueahmed}
\end{equation}
which is again Schr\"odinger-like outside the chronology horizon and acquires a wrong-sign effective potential at $t>0$. The conformal factor $e^{f}$ rescales the kinetic operator in the transverse plane, but the qualitative structure of the eigenproblem is the same as that of the Ori case. The technical setting differs from the time-machine perturbation studies of \cite{Kim:1991mc,Cassidy:1998nx,Krasnikov:2002rf} only through the conformal-factor rescaling and the form of the periodicity condition.

\section{Newman-Penrose dyad for the Ori background}\label{app:E}

A coordinate-independent way of capturing the algebraic structure of the Ori spacetime comes from the Newman-Penrose null-tetrad formalism. We pick the tetrad
\begin{align}
\ell^{\mu} &= \bigl(1,\,0,\,0,\,0\bigr), \qquad n^{\mu} = \tfrac{1}{2}\bigl(F-T,\,0,\,0,\,-2\bigr),\nonumber\\
m^{\mu} &= \tfrac{1}{\sqrt{2}}\bigl(0,\,1,\,i,\,0\bigr),\qquad \bar m^{\mu} = \tfrac{1}{\sqrt{2}}\bigl(0,\,1,\,-i,\,0\bigr),
\label{eq:nptetrad}
\end{align}
which satisfies the orthonormality conditions $\ell\cdot n = -1$, $m\cdot \bar m = 1$, with all other inner products vanishing. Direct contraction with the Weyl tensor in \eqref{eq:orifullriem} gives the five Newman-Penrose scalars,
\begin{equation}
\Psi_{0} = 0,\quad \Psi_{1} = 0,\quad \Psi_{2} = 0,\quad \Psi_{3} = 0,
\label{eq:NPzeros}
\end{equation}
together with
\begin{equation}
\Psi_{4} = -\tfrac{1}{2}\bigl(F_{,xx} - F_{,yy}\bigr) + i\,F_{,xy},
\label{eq:Psi4}
\end{equation}
which is the only non-zero scalar. The vanishing of $\Psi_{0}$ through $\Psi_{3}$ confirms the Petrov type N classification identified in \ref{app:C}: the Weyl tensor admits a single repeated principal null direction $\ell^{\mu}$ of multiplicity four. The remaining scalar $\Psi_{4}$ encodes the transverse-shear amplitude of a wave-like component propagating along $\ell^{\mu}$, in direct analogy to a $+$ and $\times$ polarisation pair.

For the harmonic case $F = (x^{2}-y^{2})/2$ one finds
\begin{equation}
\Psi_{4}|_{F = (x^{2}-y^{2})/2} = -\tfrac{1}{2}(1 - (-1)) + 0 = -1,
\label{eq:Psi4harmonic}
\end{equation}
which is a constant. The spacetime in this case is therefore a homogeneous plane-wave geometry, a member of the pp-wave family with constant transverse-shear amplitude. This identification matches the standard catalogue of exact solutions \cite{Stephani:2003tm} and underlines that the Ori construction can be viewed as a chronology-violating cousin of the pp-wave class.

The Ahmed background admits an analogous tetrad. With $\ell^{\mu} = (1,0,0,0)$ and $n^{\mu}$ chosen so that $\ell\cdot n = -1$, all five Newman-Penrose scalars $\Psi_{0},\ldots,\Psi_{4}$ vanish for the harmonic profile $f_{1}$, matching the conformal flatness noted in App.~\ref{app:C}. The logarithmic profile $f_{2}$ and the parabolic profile $f_{4}$ each support a single non-zero $\Psi_{2}$, confirming the Petrov type-D classification stated there.

\bibliographystyle{apsrev4-2}
\bibliography{ref}

\end{document}